\documentclass[11pt,a4]{article}
\pdfoutput=1 

\usepackage{jheppub} 
\usepackage[utf8]{inputenc}
\usepackage{amsfonts}
\usepackage{amsbsy}
\usepackage{url}
\usepackage{enumerate}
\usepackage[justification=raggedright]{caption}
\usepackage{hhline}
\usepackage{graphicx,xcolor}
\usepackage{amssymb,amsmath}
\usepackage{slashed}
\usepackage{hyperref}
\usepackage{braket}
\usepackage{simplewick}
\usepackage{ascmac}
\usepackage{latexsym}
\usepackage{pifont}          
\usepackage{bm}
\usepackage{comment}
\usepackage[normalem]{ulem}
\usepackage{CJKutf8}
\usepackage{cancel}
\usepackage{multicol}

\graphicspath{{./fig/}{./}{./figs/}}

\newcommand{\mr}[1]{\mathrm{#1}}

\begin{document}

\preprint{YITP-26-53, OU-HET-1307}

\title{
A simple solution to the monopole problem: \\
SU(5) GUT with symmetry breaking into special subgroup
}

\author[a]{Yu~Hamada}
\emailAdd{yu.hamada@het.phys.sci.osaka-u.ac.jp}
\affiliation[a]{Department of Physics, The University of Osaka,
Machikaneyama-cho, Toyonaka, Osaka 560- 0043, Japan}

\author[b]{and Naoki~Yamatsu}
\emailAdd{naoki.yamatsu@yukawa.kyoto-u.ac.jp}
\affiliation[b]{Yukawa Institute for Theoretical Physics, Kyoto University, Kyoto, Kyoto 606-8502, Japan}

\abstract{
Grand unified theories (GUTs) predict the overproduction of magnetic monopoles, leading to the so-called monopole problem, which is often addressed by cosmological inflation that dilutes their abundance.
However, if inflation occurs before the GUT symmetry breaking, monopoles are produced afterwards and the problem persists.
This motivates the exploration of alternative mechanisms.
We propose a simple solution based on the Langacker--Pi mechanism within an $SU(5)$ GUT framework with symmetry breaking into its special subgroup.
In particular, after the gauge symmetry is broken to the Standard Model (SM) gauge group $SU(3)_C\times SU(2)_L\times U(1)_Y$ by the vacuum expectation value of an adjoint scalar, it is further broken to $SO(3)_C\times SO(2)_L$.
This structure is naturally realized by introducing a symmetric tensor scalar, a singlet scalar, and multiple singlet fermions.
During this intermediate phase, the monopoles become connected to antimonopoles by cosmic strings, which enhances their pair annihilation and reduces their abundance.
Subsequently, the symmetry is restored to the SM gauge group.
The restoration transition can be either first-order or second-order, depending on the model parameters.
In the case of a first-order phase transition, a stochastic gravitational-wave (GW) signal is generated.
For a certain region of the parameter space, the resulting signal can lie within the sensitivity of future GW experiments.
}

\maketitle

\section{Introduction}

Grand unified theories (GUTs) provide an attractive framework
for understanding the structure of the Standard Model (SM).
In particular, they offer a unified description of the gauge
interactions and may explain the quantization of electric charge
and the origin of the fermion representations
\cite{Georgi:1974sy,Fritzsch:1974nn,Gursey:1975ki,Inoue:1977qd,Ida:1980ea,Fujimoto:1981bv}.

However, many GUTs predict the existence of topological
magnetic monopoles produced during the spontaneous symmetry
breaking of the GUT gauge group.
For example, in the minimal $SU(5)$ model
\cite{Georgi:1974sy}
the symmetry breaking
\begin{align}
SU(5)\rightarrow G_{\rm SM}:=SU(3)_C\times SU(2)_L\times U(1)_Y
\end{align}
leads to topologically stable monopoles associated with
\begin{align}
\pi_2(SU(5)/G_{\rm SM})\neq0 .
\end{align}
These monopoles arise as topologically stable solitons
associated with the spontaneous symmetry breaking of the
GUT gauge group~\cite{tHooft:1974kcl,Polyakov:1974ek}.
During the GUT phase transition they are formed via the
Kibble--Zurek mechanism~\cite{Kibble:1976sj,Zurek:1985qw}.
Their predicted abundance is many orders of magnitude larger
than the observational bounds, leading to the well-known
cosmological monopole problem~\cite{Zeldovich:1978wj,Preskill:1979zi}.

The most widely discussed solution to the monopole problem
is cosmological inflation, which exponentially dilutes the monopole
abundance to a negligible level through the expansion of the Universe
\cite{Guth:1980zm,Linde:1981mu}.
While this mechanism provides an elegant cosmological
resolution, it leads to a bound on the inflation scale, which is required to lie below the GUT-breaking scale.
This bound becomes even more severe in scenarios where magnetic monopoles are produced during a later stage of symmetry breaking.
A typical example arises in models with multiple symmetry-breaking stages and intermediate scales, such as
$SO(10)\rightarrow SU(4)_C\times SU(2)_L\times SU(2)_R\rightarrow G_\mr{SM}$,
where the intermediate gauge group corresponds to the Pati--Salam symmetry \cite{Pati:1974yy}.
Such symmetry-breaking chains have been widely studied in the literature \cite{Deshpande:1992au,Deshpande:1992em,Bertolini:2009qj,Altarelli:2013aqa,Ferrari:2018rey,Chakrabortty:2019fov,Abe:2021byq,Okada:2021qmi}.
In such cases, the corresponding phase transition may occur after inflation, thereby reintroducing the monopole problem.
Therefore, it is also interesting to explore alternative mechanisms that can eliminate monopoles without invoking inflation.

\begin{figure}
    \centering
    \includegraphics[width=0.55\linewidth]{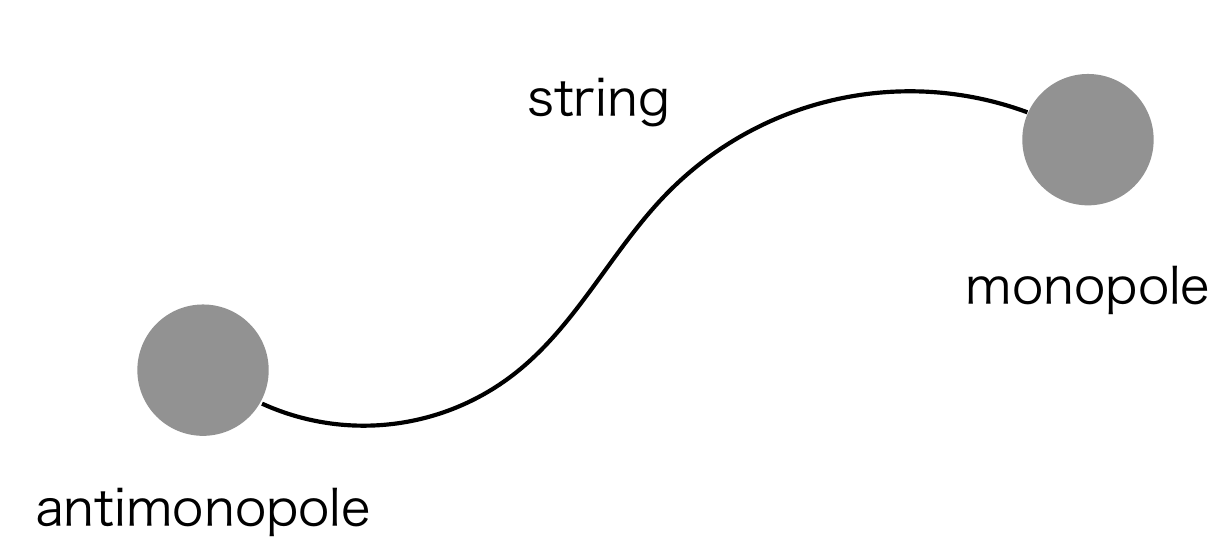}
    \caption{Illustration of a segment consisting of monopole, antimonopole and cosmic string,
    which appears in the Langacker-Pi mechanism.}
    \label{fig:segment}
\end{figure}

One such possibility was proposed by Langacker and Pi
\cite{Langacker:1980kd}.
In the Langacker--Pi mechanism, monopoles are first produced
during the initial symmetry breaking of the GUT gauge group.
At a later stage of the cosmological evolution,
a $U(1)$ gauge symmetry is spontaneously broken,
which leads to the formation of cosmic strings
\cite{Kibble:1976sj,Vilenkin:2000jqa}.
These strings attach to monopoles and antimonopoles,
forming string--monopole systems.
(See Fig.~\ref{fig:segment}.)
As the strings shrink due to their tension,
the monopole--antimonopole pairs are pulled together
and eventually annihilate.
In this way the monopole abundance can be efficiently
reduced through the subsequent phase transition.
For other scenarios to erase monopoles, see Refs.~\cite{Dvali:1997sa,Pogosian:1999zi,Brush:2015vda,Bachmaier:2023zmq,Bachmaier:2023wzz,Bachmaier:2026bgw}.

In order to realize the Langacker--Pi mechanism in GUT
models, it is necessary to consider symmetry-breaking
patterns that lead to the formation of cosmic strings
at a later stage of the cosmological evolution.
One simple possibility is that the $U(1)_\mr{EM}$ or $U(1)_Y$ gauge symmetry is spontaneously broken around the electroweak scale due to some extension of the SM Higgs sector~\cite{Farris:1991rg}.
Nevertheless, it is not clear if such a scenario works efficiently~\cite{Holman:1992xs,Gates:1992gd}.
An alternative and even simpler possibility arises when the GUT gauge group
is broken into a \emph{special subgroup}
\cite{Dynkin:1957ek,Dynkin:1957um,Slansky:1981yr,Cahn:1985wk,Yamatsu:2015npn}.\footnote{
It has been pointed out~\cite{Farris:1991rg} that antisymmetric tensor scalars in $SU(5)$ may, in principle, allow for breaking of the $U(1)_Y$ factor. 
However, realizing a vacuum that breaks only $U(1)_Y$ while preserving $SU(3)_C\times SU(2)_L$ requires a nontrivial alignment of the VEV in group space and may not be generic.}
Unlike the so-called regular subgroups, which are obtained
by deleting nodes from the Dynkin diagram of the original
group, special subgroups correspond to nontrivial embeddings
of smaller groups into the GUT gauge group.
Symmetry breaking into such subgroups can therefore lead
to vacuum manifolds whose topological properties differ
significantly from those arising in conventional
symmetry-breaking patterns.
In particular, when the gauge symmetry temporarily passes
through a phase associated with a special subgroup,
the vacuum manifold may change so that
stable monopoles are no longer supported
and eventually disappear due to the Langacker--Pi mechanism.
Therefore the symmetry breaking into special subgroups may naturally realize the Langacker--Pi mechanism.
Although this observation is especially relevant in the
cosmological context, the properties of symmetry breaking
into special subgroups themselves have been studied more
generally in various gauge-theory frameworks
\cite{Pati:1975ca,Pati:1980sb,Yamatsu:2017mei,Yamatsu:2017woc,Yamatsu:2018tnv,Kugo:2019isl,Kugo:2019wge}.

In this paper we investigate the realization of the
Langacker--Pi mechanism in a scenario where the symmetry
breaking temporarily proceeds through a special subgroup.
In particular, we consider an $SU(5)$ model containing a
symmetric tensor scalar field in the $\mathbf{15}$
representation.
The vacuum expectation value (VEV) of this field can break
$SU(5)$ into the special subgroup $SO(5)$.
The temporary appearance of this phase plays an important
role in the cosmological evolution of the monopole system.

A simple realization of this idea can be obtained
in an $SU(5)$ model with an extended scalar sector.
The symmetry-breaking sequence considered in this work
can be schematically written as
\begin{align}
SU(5)
\;\rightarrow\;
G_{\rm SM}
\;\rightarrow \;
SO(3)_C\times SO(2)_L
\; \rightarrow \;
G_{\rm SM}. \label{eq:sequence}
\end{align}
The first step of this sequence is realized by the adjoint
scalar field $\Phi_{\bf 24}$, whose VEV
breaks the $SU(5)$ gauge symmetry into the SM gauge group $G_{\rm SM}=SU(3)_C\times SU(2)_L\times U(1)_Y$,
producing magnetic monopoles.
In the present framework this breaking occurs at the
GUT scale and sets the stage for the subsequent
cosmological evolution of the scalar sector.

\begin{figure}
    \centering
    \includegraphics[width=0.5\linewidth]{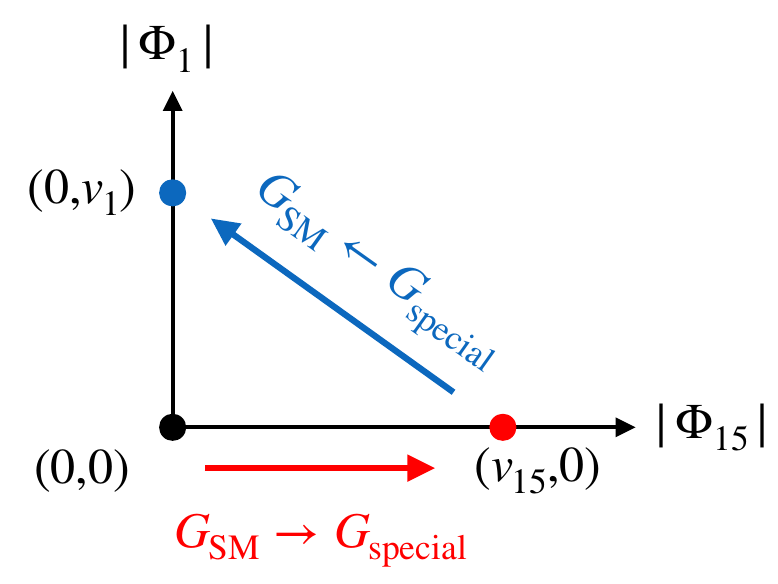}
    \caption{Schematic picture of the symmetry breakings in our scenario.
    The horizontal and diagonal arrows in the figure correspond to the second and third arrows in Eq.~\eqref{eq:sequence}, respectively.
    $G_\mr{special}$ is the special subgroup of $G_\mr{SM}$, $G_\mr{special} := SO(3)_C\times SO(2)_L$.
    We ignore the direction of the adjoint scalar $\Phi_\textbf{24}$ because it is effectively decoupled due to the hierarchical VEVs.
    $\Phi_{\bf 15}$ and  $\Phi_{\bf 1}$ denote the scalar in the $\mathbf{15}$ representation and the singlet scalar, respectively.}
    \label{fig:two-step}
\end{figure}

In the intermediate phase the vacuum is aligned along
an $SO(5)$ direction induced by the symmetric tensor scalar.
In the presence of the previously generated
$G_{\rm SM}$ structure, the effective unbroken symmetry
is reduced to the subgroup
$SO(3)_C\times SO(2)_L$.
This transition is indicated by the red horizontal arrow in Fig.~\ref{fig:two-step}.
Due to the absence of the $U(1)_Y$ factor,
the topology of the corresponding vacuum manifold does not allow stable magnetic monopoles,
which are instead connected by the $U(1)_Y$ strings
and disappear by the Langacker--Pi mechanism.
Afterwards the symmetry is restored back to $G_\mr{SM}$,
which occurs as the VEV is transferred from the symmetric tensor scalar to the singlet scalar,
shown by the blue diagonal arrow in Fig.~\ref{fig:two-step}.
We find that this transition can be a first-order phase transition (FOPT) accompanied by bubble nucleation in a large parameter space of the model, 
so that the dynamics of bubbles during the phase transition may lead to a stochastic gravitational wave (GW)
background~\cite{Turner:1990rc,Kosowsky:1991ua,Kosowsky:1992vn,Kosowsky:1992rz,Kamionkowski:1993fg}, providing a possible observational signature.

The main purpose of the present work is to clarify whether
such a symmetry-breaking sequence can be realized within a
relatively simple scalar sector.
We therefore analyze the scalar potential containing the
adjoint, symmetric tensor, fundamental, and singlet scalar
fields and study the resulting vacuum structure.
Special attention is paid to the role of the symmetric
tensor scalar in realizing the intermediate phase
associated with the special subgroup.

This paper is organized as follows.
In Sec.~2 we introduce our $SU(5)$ model.
In Sec.~3 we analyze the vacuum structure of the scalar potential
and derive the parameter conditions required to realize the desired
pattern of spontaneous symmetry breaking.
In Sec.~4 we describe the cosmological scenario based on this
symmetry-breaking sequence and discuss how the monopole problem
can be addressed through the Langacker--Pi mechanism.
In Sec.~5 we investigate the FOPT associated
with the intermediate symmetry restoration and evaluate the resulting
GW spectrum.
Section~6 is devoted to discussion and conclusions.
The vacuum structures for the individual scalar fields used in the
analysis are summarized in Appendix~A.

\section{$SU(5)$ Model}

We consider an $SU(5)$ GUT \cite{Georgi:1974sy}
with additional scalar fields
in a symmetric tensor representation,
together with a singlet scalar
and singlet fermions.
Compared with the minimal $SU(5)$ model, 
we introduce an additional scalar field
$\Phi_{\bf 15}$ transforming as the
symmetric rank-two tensor representation
of $SU(5)$.
The symmetric tensor scalar plays an essential role
in realizing symmetry breaking into the special subgroup
$SO(5)$.
We also introduce a singlet scalar $\Phi_{\bf 1}$ and $N_f$ singlet fermion $\Psi_{\bf 1}^{(i)}$
to realize the appropriate symmetry breaking pattern.
The field contents of the model are summarized in
Table~\ref{tab:fields}.
The $SU(5)$ indices are denoted by
\(
a,b=1,2,...,5
\),
while the non-singlet fermion generation indices are denoted by
\(
\alpha,\beta(=1,2,3)
\), and the singlet fermion indices are denoted by \(i, j (=1,2,...,N_f)\).

\begin{table}[t]
\centering
\begin{tabular}{|c|c|c|c|c|c|c|c|c|c|}\hline
 4D field
 &{${A}_\mu$}
 &{${\Psi}_{\bf 10}^{(\alpha)}$}
 &{${\Psi}_{\bf \overline{5}}^{(\alpha)}$}
 &{${\Psi}_{\bf 1}^{(i)}$}
 &{${\Phi}_{\bf 24}$}
 &{${\Phi}_{\bf 5}$}
 &{${\Phi}_{\bf 15}$}
 &{${\Phi}_{\bf 1}$}
 \\\hline
 $SU(5)$
 &{${\bf 24}$}
 &{${\bf 10}$}
 &{${\bf \overline{5}}$}
 &{${\bf 1}$}
 &{${\bf 24}$}
 &{${\bf 5}$}
 &{${\bf 15}$}
 &{${\bf 1}$}
\\\hline
 $SL(2,\mathbb{C})$
 &$(1/2,1/2)$
 &$(1/2,0)$
 &$(1/2,0)$
 &$(1/2,0)$
 &$(0,0)$
 &$(0,0)$
 &$(0,0)$
 &$(0,0)$
 \\\hline
\end{tabular}
\caption{The field contents of the model are summarized.
The generation indices are
$\alpha=1,2,3$.
The index $i$ labels singlet fermions. $i=1,2,...,N_f$.}
\label{tab:fields}
\end{table}

The Lagrangian of the model consists of the gauge,
fermion, and scalar sectors,
\begin{align}
\mathcal{L}
=
\mathcal{L}_{\rm gauge}
+
\mathcal{L}_{\rm fermion}
+
\mathcal{L}_{\rm scalar}
+
\mathcal{L}_{\rm Yukawa+mass}
-
V .
\end{align}
The gauge kinetic term is given by
\begin{align}
\mathcal{L}_{\rm gauge}
=
-\frac14
F^A_{\mu\nu}F^{A\mu\nu},
\end{align}
with
\begin{align}
F^A_{\mu\nu}
=
\partial_\mu A^A_\nu
-
\partial_\nu A^A_\mu
+
g f^{ABC}A^B_\mu A^C_\nu ,
\qquad
(A_\mu)_a^{\ b} = A_\mu^A (T^A)_a^{\ b},
\end{align}
where 
$A=1,\dots,24$, 
${\rm Tr}(T^A T^B)=\delta^{AB}/2$, 
$(T^A)_a^{\ b}$ are the generators of $SU(5)$ in the fundamental representation and
$g$ is the gauge coupling constant.

The fermion kinetic terms are given as
\begin{align}
\mathcal{L}_{\rm fermion}
=
&i
\overline{\Psi}^{(\alpha)}_{\bf 10}
\gamma^\mu D_\mu
\Psi^{(\alpha)}_{\bf 10}
+
i
\overline{\Psi}^{(\alpha)}_{\bf \overline5}
\gamma^\mu D_\mu
\Psi^{(\alpha)}_{\bf \overline5}
+
i
\overline{\Psi}^{(i)}_{\bf 1}
\gamma^\mu \partial_\mu
\Psi^{(i)}_{\bf 1},
\end{align}
where 
 $\Psi_{\bf \overline5}$, 
 $\Psi_{\bf 10}$, and
 $\Psi_{\bf 1}$
 are 
 the fermion fields in the $\mathbf{\bar 5}$, $\mathbf{10}$, and singlet representations, respectively.
 Since  $\mathbf{10}$ is the antisymmetric representation, we have
 \begin{align}
(\Psi_{\bf 10}^{(\alpha)})_{ab} = -(\Psi_{\bf 10}^{(\alpha)})_{ba}.
\end{align}
Their covariant derivatives are defined as
\begin{align}
(D_\mu \Psi_{\bf \overline5}^{(\alpha)})^a
&=
(\partial_\mu \Psi_{\bf \overline5}^{(\alpha)})^a
+
i g (A_\mu)_b^{\ a}(\Psi_{\bf \overline5}^{(\alpha)})^b ,
\nonumber\\
(D_\mu \Psi_{\bf 10}^{(\alpha)})_{ab}
&=
\partial_\mu (\Psi_{\bf 10}^{(\alpha)})_{ab}
-
i g (A_\mu)_a^{\ c}(\Psi_{\bf 10}^{(\alpha)})_{cb}
-
i g (A_\mu)_b^{\ c}(\Psi_{\bf 10}^{(\alpha)})_{ac}.
\end{align}

The scalar kinetic terms are given by 
\begin{align}
\mathcal{L}_{\rm scalar}
=
(D_\mu\Phi_{\bf 24})^\dagger(D^\mu\Phi_{\bf 24})
+
(D_\mu\Phi_{\bf 5})^\dagger(D^\mu\Phi_{\bf 5})
+
(D_\mu\Phi_{\bf 15})^\dagger(D^\mu\Phi_{\bf 15})
+
\frac12(\partial_\mu\Phi_{\bf 1})^2 ,
\end{align}
where $\Phi_{\bf 24}$, $\Phi_{\bf 5}$, $\Phi_{\bf 15}$, and $\Phi_{\bf 1}$ are in the adjoint, fundamental, symmetric rank-two tensor, and singlet representations, respectively.
With the explicit $SU(5)$ indices, they are often written as
\begin{align}
(\Phi_{\bf 24})_a^{\ b}, 
\qquad
(\Phi_{\bf 5})_a ,
\qquad 
(\Phi_{\bf 15})_{ab} \, ,
\end{align}
which satisfy
\begin{align}
&(\Phi_{\bf 24})_a^{\ a}=0 ,
\quad \text{(traceless Hermitian)} 
\nonumber\\
&(\Phi_{\bf 15})_{ab}=(\Phi_{\bf 15})_{ba}. \quad \text{(symmetric tensor)} 
\end{align}
They transform under an $SU(5)$ gauge transformation $U$ as
\begin{align}
(\Phi_{\bf 24})_a^{\ b}
&\rightarrow
U_a^{\ c}(\Phi_{\bf 24})_c^{\ d}(U^\dagger)_d^{\ b},
\qquad (\Phi_{\bf 24}\rightarrow U\Phi_{\bf 24}U^\dagger ) 
\nonumber\\
(\Phi_{\bf 5})_a
&\rightarrow
U_a^{\ b}(\Phi_{\bf 5})_b ,
\qquad\qquad  (\Phi_{\bf 5}\rightarrow U\Phi_{\bf 5} ) 
\nonumber\\
(\Phi_{\bf 15})_{ab}
&\rightarrow
U_a^{\ c} U_b^{\ d} (\Phi_{\bf 15})_{cd}, 
\qquad (\Phi_{\bf 15} \rightarrow U \Phi_{\bf 15} U^T ),
\end{align}
where the transformation matrix is defined as
\begin{align}
U_a^{\ b}
=
\left[\exp\!\left(i\alpha^A T^A\right)\right]_a^{\ b},
\qquad A=1,\dots,24 .
\end{align}
Their covariant derivatives are given as
\begin{align}
(D_\mu\Phi_{\bf 24})_a^{\ b}
&=
\partial_\mu(\Phi_{\bf 24})_a^{\ b}
-i g (A_\mu)_a^{\ c}(\Phi_{\bf 24})_c^{\ b}
+i g (\Phi_{\bf 24})_a^{\ c}(A_\mu)_c^{\ b}.
\nonumber\\
(D_\mu\Phi_{\bf 5})_a
&=
\partial_\mu(\Phi_{\bf 5})_a
-i g (A_\mu)_a^{\ b}(\Phi_{\bf 5})_b .
\nonumber\\
(D_\mu \Phi_{\bf 15})_{ab}
&=
\partial_\mu (\Phi_{\bf 15})_{ab}
- i g (A_\mu)_a^{\ c}(\Phi_{\bf 15})_{cb}
- i g (A_\mu)_b^{\ c}(\Phi_{\bf 15})_{ac}.
\end{align}

The Yukawa interactions are given by
\begin{align}
\mathcal{L}_{\rm Yukawa+mass}
&=
(y_{u})_{\alpha\beta}\,
(\Psi_{\bf 10}^{(\alpha)})_{ab}
(\Psi_{\bf 10}^{(\beta)})_{cd}
(\Phi_{\bf 5})_e\,
\epsilon^{abcde}
+
(y_{d})_{\alpha\beta}\,
(\Psi_{\bf 10}^{(\alpha)})_{ab}
(\Psi_{\bf \overline5}^{(\beta)})^a
(\Phi_{\bf 5}^{\dagger })^b
\nonumber\\
&\quad
+
(y)_{ij}\,
\Psi_{\bf 1}^{(i)}
\Psi_{\bf 1}^{(j)}
\Phi_{\bf 1}
+
(y_{15})_{\alpha\beta}\,
(\Phi_{\bf 15})_{ab}
(\Psi_{\bf \overline5}^{(\alpha)})^a
(\Psi_{\bf \overline5}^{(\beta)})^b
\nonumber\\
&\quad
+
(y_{5})_{i\alpha}\,
\Psi_{\bf 1}^{(i)}
(\Psi_{\bf \overline 5}^{(\alpha)})^{a}
(\Phi_{\bf 5})_a
+{\rm h.c.}
\nonumber \\
&\quad 
-\frac12
(M_{\bf 1})_{ij}
\Psi_{\bf 1}^{(i)}
\Psi_{\bf 1}^{(j)} .
\label{eq:yukawa}
\end{align}
Here $\alpha,\beta=1,2,3$ denote the generation indices of the
$SU(5)$ matter multiplets, while $i,j=1,\dots,N_f$ label the flavors
of the singlet fermions.
The matrices $(y_x)_{\alpha\beta}$ $(x=u,d, 15,)$, $(y_5)_{i\alpha}$ and $(y)_{ij}$ represent Yukawa coupling
matrices in flavor space, and $M_{\bf 1}$ is the Majorana mass matrix
of the singlet fermions.
Hereafter we take $M_{\bf 1}=0$ in our scenario,
which can be naturally realized by imposing a global symmetry corresponding to the fermion number of $\Psi_\textbf{1}^{(i)}$'s.

The renormalizable scalar potential consists of the self-interaction
terms of each scalar multiplet and the mixing terms among them.
It can be written as
\begin{align}
V(\Phi_{\bf 24},\Phi_{\bf 5},\Phi_{\bf 15},\Phi_{\bf 1})
=
\sum_{{\bf x}={\bf 24},{\bf 5},{\bf 15},{\bf 1}}V_{\bf x}(\Phi_{\bf x})
+
V_{\rm mix}(\Phi_{\bf 24},\Phi_{\bf 5},\Phi_{\bf 15},\Phi_{\bf 1}) .
\label{Eq:Potential_Total}
\end{align}
Here $V_{\bf x}$ denotes each self-interaction potential for the scalar multiplet
$\Phi_{\bf x}$
with ${\bf x}={\bf 24},{\bf 5},{\bf 15},{\bf 1}$,
while $V_{\rm mix}$ contains the interaction terms
among different scalar fields.
The individual potentials are given by
\begin{align}
V_{\bf 24}(\Phi_{\bf 24})
&=
-\frac12 \mu_{\bf 24}^2
{\rm Tr}(\Phi_{\bf 24}^2)
+\frac13 \kappa_{\bf 24}\,
{\rm Tr}(\Phi_{\bf 24}^3)
+
\frac14 \lambda_{{\bf 24}1}
\left[{\rm Tr}(\Phi_{\bf 24}^2)\right]^2
+
\frac14 \lambda_{{\bf 24}2}
{\rm Tr}(\Phi_{\bf 24}^4),
\label{Eq:Potential_V24}
\\
V_{\bf 5}(\Phi_{\bf 5})
&=
-\frac12\mu_{\bf 5}^2
(\Phi_{\bf 5}^\dagger \Phi_{\bf 5})
+
\frac14\lambda_{\bf 5}
(\Phi_{\bf 5}^\dagger \Phi_{\bf 5})^2,
\label{Eq:Potential_V5}
\\
V_{\bf 15}(\Phi_{\bf 15})
&=
-\frac12 \mu_{\bf 15}^2
{\rm Tr}(\Phi_{\bf 15}^\dagger\Phi_{\bf 15})
+
\frac14 \lambda_{{\bf 15}1}
[{\rm Tr}(\Phi_{\bf 15}^\dagger\Phi_{\bf 15})]^2
+
\frac14 \lambda_{{\bf 15}2}
{\rm Tr}(\Phi_{\bf 15}^\dagger\Phi_{\bf 15}
\Phi_{\bf 15}^\dagger\Phi_{\bf 15}),
\label{Eq:Potential_V15}
\\
V_{\bf 1}(\Phi_{\bf 1})
&=
-\frac12\mu_{\bf 1}^2 \Phi_{\bf 1}^2
+
\frac13 \kappa_{\bf 1}\,
\Phi_{\bf 1}^3
+
\frac14\lambda_{\bf 1}\Phi_{\bf 1}^4 .
\label{Eq:Potential_V1}
\end{align}
Here $\lambda_{\bf x}$, $\kappa_{\bf x}$, $\mu_{\bf x}$ 
denote quartic, cubic, quadratic scalar coupling constants, respectively.
The mixing interactions among different scalar multiplets are
\begin{align}
V_{\rm mix}(\Phi_{\bf 24},\Phi_{\bf 5},\Phi_{\bf 15},\Phi_{\bf 1})
&=
 V_{\rm mix}{}_{;{\bf 24},{\bf  5}}(\Phi_{\bf 24},\Phi_{\bf  5})
+V_{\rm mix}{}_{;{\bf 24},{\bf 15}}(\Phi_{\bf 24},\Phi_{\bf 15})
+V_{\rm mix}{}_{;{\bf 24},{\bf  1}}(\Phi_{\bf 24},\Phi_{\bf  1}) 
\nonumber\\
&\quad
+V_{\rm mix}{}_{;{\bf 15},{\bf  1}}(\Phi_{\bf 15},\Phi_{\bf  1})
+V_{\rm mix}{}_{;{\bf 15},{\bf  5}}(\Phi_{\bf 15},\Phi_{\bf  5})
+V_{\rm mix}{}_{;{\bf  1},{\bf  5}}(\Phi_{\bf  1},\Phi_{\bf  5})
\nonumber\\
&\quad
+V_{\rm mix}{}_{;{\bf 24},{\bf 15},{\bf 1}}(\Phi_{\bf 24},\Phi_{\bf 15},\Phi_{\bf 1})
+V_{\rm mix}{}_{;{\bf 24},{\bf 1},{\bf 5}}(\Phi_{\bf 24},\Phi_{\bf 5},\Phi_{\bf 5})
\nonumber\\
&\quad
+V_{\rm mix}{}_{;{\bf 24},{\bf 15},{\bf 5}}(\Phi_{\bf 24},\Phi_{\bf 15},\Phi_{\bf 5})
+V_{\rm mix}{}_{;{\bf 15},{\bf 1},{\bf 5}}(\Phi_{\bf 15},\Phi_{\bf 1},\Phi_{\bf 5}),
\label{Eq:Potential_Vmix}
\end{align}
where the mixing terms composed of two scalar fields are
\begin{align}
V_{\rm mix}{}_{;{\bf 24},{\bf  5}}(\Phi_{\bf 24},\Phi_{\bf  5})
&=
\frac14\lambda_{{\bf 24},{\bf 5}1}
(\Phi_{\bf 5}^\dagger \Phi_{\bf 5})
{\rm Tr}(\Phi_{\bf 24}^2)
+
\frac14\lambda_{{\bf 24},{\bf 5}2}
\Phi_{\bf 5}^\dagger \Phi_{\bf 24}^2 \Phi_{\bf 5}
\nonumber\\
&\quad
+
\frac12\kappa_{{\bf 24},{\bf 5}}\,
\Phi_{\bf 5}^\dagger
\Phi_{\bf 24}
\Phi_{\bf 5}
\label{Eq:Potential_Vmix-24+5}
,\\
V_{\rm mix}{}_{;{\bf 24},{\bf 15}}(\Phi_{\bf 24},\Phi_{\bf 15})
&=
\frac14\lambda_{{\bf 24},{\bf 15}1}
{\rm Tr}(\Phi_{\bf 24}^2)
{\rm Tr}(\Phi_{\bf 15}^\dagger\Phi_{\bf 15})
+
\frac14\lambda_{{\bf 24},{\bf 15}2}
{\rm Tr}(\Phi_{\bf 15}^\dagger \Phi_{\bf 24}^2 \Phi_{\bf 15})
\nonumber\\
&\quad
+
\frac12\kappa_{{\bf 24},{\bf 15}}\,
{\rm Tr}(\Phi_{\bf 15}^\dagger
\Phi_{\bf 24}
\Phi_{\bf 15})
\label{Eq:Potential_Vmix-24+15}
,\\
V_{\rm mix}{}_{;{\bf 24},{\bf  1}}(\Phi_{\bf 24},\Phi_{\bf  1})
&=
\frac12\lambda_{{\bf 24},{\bf 1}}
\Phi_{\bf 1}^2
{\rm Tr}(\Phi_{\bf 24}^2)
+
\frac12\kappa_{{\bf 24},{\bf 1}}\,
\Phi_{\bf 1}
{\rm Tr}(\Phi_{\bf 24}^2)
\label{Eq:Potential_Vmix-24+1}
,\\
V_{\rm mix}{}_{;{\bf 15},{\bf  1}}(\Phi_{\bf 15},\Phi_{\bf  1})
&=
\frac14\lambda_{{\bf 15},{\bf 1}}
\Phi_{\bf 1}^2
{\rm Tr}(\Phi_{\bf 15}^\dagger\Phi_{\bf 15})
+
\frac12\kappa_{{\bf 15},{\bf 1}}\,
\Phi_{\bf 1}
{\rm Tr}(\Phi_{\bf 15}^\dagger
\Phi_{\bf 15})
\label{Eq:Potential_Vmix-15+1}
,\\
V_{\rm mix}{}_{;{\bf 15},{\bf  5}}(\Phi_{\bf 15},\Phi_{\bf  5})
&=
\frac14\lambda_{{\bf 15},{\bf 5}}
(\Phi_{\bf 5}^\dagger \Phi_{\bf 5})
{\rm Tr}(\Phi_{\bf 15}^\dagger\Phi_{\bf 15})
+
\frac12\kappa_{{\bf 15},{\bf 5}}\,
\Phi_{\bf 5}^\dagger
\Phi_{\bf 15}
\Phi_{\bf 5}^*
+{\rm h.c.}
\label{Eq:Potential_Vmix-15+5}
,\\
V_{\rm mix}{}_{;{\bf  1},{\bf  5}}(\Phi_{\bf  1},\Phi_{\bf  5})
&=
\frac12\lambda_{{\bf 1},{\bf 5}}
\Phi_{\bf 1}^2
(\Phi_{\bf 5}^\dagger \Phi_{\bf 5})
+
\frac12\kappa_{{\bf 1},{\bf 5}}\,
\Phi_{\bf 1}
\Phi_{\bf 5}^\dagger
\Phi_{\bf 5},
\label{Eq:Potential_Vmix-5+1}
\end{align}
and the mixing terms composed of three scalar fields are
\begin{align}
V_{\rm mix}{}_{;{\bf 24},{\bf 15},{\bf 1}}(\Phi_{\bf 24},\Phi_{\bf 15},\Phi_{\bf 1})
&=
\frac12\lambda_{{\bf 24},{\bf 15},{\bf 1}}\,
\Phi_{\bf 1}
{\rm Tr}(\Phi_{\bf 15}^\dagger
\Phi_{\bf 24}
\Phi_{\bf 15})
\label{Eq:Potential_Vmix-24+15+1}
,\\
V_{\rm mix}{}_{;{\bf 24},{\bf 1},{\bf 5}}(\Phi_{\bf 24},\Phi_{\bf 5},\Phi_{\bf 5})
&=
\frac12\lambda_{{\bf 24},{\bf 1},{\bf 5}}\,
\Phi_{\bf 1}
\Phi_{\bf 5}^\dagger
\Phi_{\bf 24}
\Phi_{\bf 5}
\label{Eq:Potential_Vmix-24+1+5}
,\\
V_{\rm mix}{}_{;{\bf 24},{\bf 15},{\bf 5}}(\Phi_{\bf 24},\Phi_{\bf 15},\Phi_{\bf 5})
&=
\frac12\lambda_{{\bf 24},{\bf 15},{\bf 5}}\,
\Phi_{\bf 5}^\dagger
\Phi_{\bf 24}
\Phi_{\bf 15}
\Phi_{\bf 5}^*
+{\rm h.c.}
\label{Eq:Potential_Vmix-24+15+5}
,\\
V_{\rm mix}{}_{;{\bf 15},{\bf 1},{\bf 5}}(\Phi_{\bf 15},\Phi_{\bf 1},\Phi_{\bf 5})
&=
\frac12\lambda_{{\bf 15},{\bf 1},{\bf 5}}\,
\Phi_{\bf 1}
\Phi_{\bf 5}^\dagger
\Phi_{\bf 15}
\Phi_{\bf 5}^*
+{\rm h.c.}
\label{Eq:Potential_Vmix-15+1+5}
\end{align}


\section{Vacuum structure and parameter constraints}

The VEVs of the scalar fields
are determined by minimizing the scalar potential given in
Eq.~(\ref{Eq:Potential_Total}).
Denoting the relevant VEVs collectively as
\begin{align}
V(\{v_{\bf x}\}) := V(v_{\bf 24},v_{\bf 5},v_\textbf{15},v_{\bf 1}),
\end{align}
the stationary configurations are obtained from
\begin{align}
\frac{\partial V(\{v_{\bf x}\})}{\partial v_{\bf x}}=0 .
\end{align}
These conditions determine candidate extrema of the potential,
and the physical vacuum is identified by selecting the
configuration with the lowest energy.
The corresponding symmetry-breaking pattern is determined by
the set of nonvanishing VEVs.

The scalar potential in our model Eq.~(\ref{Eq:Potential_Total}) contains four scalar
multiplets $\Phi_{\bf x}$.
The vacuum structure associated with each individual scalar
potential $V_{\bf x}(\Phi_{\bf x})$ has been studied previously,
most notably in Ref.~\cite{Li:1973mq}.
For completeness, a brief summary of these results is given in
Appendix~\ref{Appendix:Vacuum}.
Here we instead focus on the vacuum structure of the
full scalar potential containing all four scalar multiplets.
In particular, we analyze how the mixing interactions among
different scalar fields affect the realization of the vacuum.

Since the scalar sector contains fields whose 
VEVs are expected to appear at very different energy
scales, it is convenient to analyze the scalar potential by
exploiting the hierarchy among the VEVs.
The adjoint scalar $\Phi_{\bf 24}$ is responsible for the breaking
of the GUT gauge symmetry
\begin{align}
SU(5)\rightarrow G_{\rm SM},
\end{align}
and its VEV is therefore assumed to be of order the
GUT scale, as given in Eq.~(\ref{Eq:VEV-24-5-321}),
satisfying
\begin{align}
v_{\bf 24} \gg v_\textbf{15},\, v_{\bf 1},\, v_{\bf 5} .
\end{align}
Under this hierarchy the adjoint field can be treated as a
background field when studying the vacuum structure of the
remaining scalar multiplets.
Through the mixing terms such as $V_{\rm mix}{}_{;{\bf 24},{\bf 15}}$,
the VEV $v_\textbf{24}$ effectively induces self-interaction terms like $\lambda_{{\bf 24},{\bf15}1}\, v_{\bf 24}^2 \mr{Tr}(\Phi_{\bf 15}^\dagger \Phi_{\bf15})$.
Although some of them can be absorbed into the existing self-interaction terms,
we simply assume here all of them to be negligibly small.

Meanwhile,
the scalar multiplet $\Phi_{\bf 5}$ contains the SM Higgs doublet,
and its VEV is assumed to lie at the electroweak scale,
\begin{align}
v_{\bf 5} \sim 10^2\ \mathrm{GeV},
\end{align}
which is much smaller than the other symmetry-breaking scales:
$v_{\bf 24} \gg v_\textbf{15}, v_{\bf 1} \gg v_{\bf 5}$.
Consequently, in the first approximation we neglect the effects
of $\Phi_{\bf 5}$ when analyzing the structure of the
higher-scale vacuum.

Under these assumptions the dominant effects in the intermediate
stage of symmetry breaking arise from the interactions between
the symmetric tensor scalar $\Phi_{\bf 15}$ and the singlet
scalar $\Phi_{\bf 1}$.
We therefore first analyze the vacuum structure in the
$(\Phi_{\bf 15},\Phi_{\bf 1})$ sector.
Keeping only the terms involving these fields in
Eq.~(\ref{Eq:Potential_Total}), the scalar potential reduces to
\begin{align}
V(\Phi_{\bf 15},\Phi_{\bf 1})
:=
V_{\bf 15}(\Phi_{\bf 15})
+
V_{\bf 1}(\Phi_{\bf 1})
+
V_{\rm mix}{}_{;{\bf 15},{\bf  1}}(\Phi_{\bf 15},\Phi_{\bf  1}) .
\label{Eq:Potential_15+1}
\end{align}
We first focus on the individual potentials $V_{\bf 15}(\Phi_{\bf 15})$ and $V_{\bf 1}(\Phi_{\bf 1})$ ignoring the mixing part $V_{\rm mix}{}_{;{\bf 15},{\bf  1}}(\Phi_{\bf 15},\Phi_{\bf  1})$.

In order to realize the Langacker-Pi mechanism, the symmetric tensor scalar $\Phi_{\bf 15}$ should develop a non-trivial VEV at an intermediate stage.
A simple way to achieve this is to assume the potential $V_{\bf 15}(\Phi_{\bf 15})$ to have a non-trivial vacuum in the absence of the mixing interaction
since, at that stage, the singlet scalar $\Phi_{\bf 1}$ is expected to have a negligible VEV due to the thermal potential as discussed in the next section.  
From the analysis summarized in
Appendix~\ref{Appendix:Vacuum-Tensor},
the realization of this (transient) vacuum requires the parameter conditions
\begin{align}
\mu_{\bf 15}^2>0,
\qquad
\lambda_{{\bf 15}1}>0,
\qquad
\lambda_{{\bf 15}2}>0 ,
\end{align}
which ensure the stability of the potential and selects the
$SU(5)\rightarrow SO(5)$ breaking direction.
In the presence of a non-vanishing VEV of the adjoint scalar
$\Phi_{\bf 24}$, the $SU(5)$ symmetry is already broken down to
$G_{\rm SM}$, so that the VEV of $\Phi_{\bf 15}$ effectively
induces the symmetry breaking
$G_{\rm SM} \rightarrow SO(3)_C \times SO(2)_L$.

Similarly, $V_{\bf 1}(\Phi_{\bf 1})$ should also have a non-trivial minimum without the mixing terms
because the absolute minimum should be $\Phi_{\bf 1}\neq 0$ and $\Phi_{\bf 15}=0$ at zero temperature.
From Appendix~\ref{Appendix:Vacuum-Singlet},
this requires
\begin{align}
\mu_{\bf 1}^2>0,
\qquad
\lambda_{\bf 1}>0 .
\end{align}
Although the cubic coupling $\kappa_{\bf 1}$ is in principle arbitrary,
it does not play an essential role in the present discussion.
For simplicity, we therefore set
\begin{align}
\kappa_{\bf 1}=0 .
\end{align}

Next we take into account the mixing
interactions given in Eq.~(\ref{Eq:Potential_Vmix-15+1}).
The cubic coupling $\kappa_{{\bf 15},{\bf 1}}$
is also not crucial for the qualitative features of the vacuum
structure.
In order to simplify the analysis, we neglect it by
setting\footnote{
Setting $\kappa_{\textbf{1}}=\kappa_\textbf{15,1}=0$ 
enhances a $\mathbb{Z}_2$ symmetry $\Phi_\textbf{1} \to - \Phi_\textbf{1}$,
leading to a domain wall production when it gets the VEV
and causing the so-called domain wall problem.
Nevertheless, this can be easily avoided by introducing tiny $\kappa_{\textbf{1}}$ and $\kappa_\textbf{15,1}$ as small as $v_\textbf{1}^3/M_\mathrm{pl}^2$,
which are negligible for the vacuum structure and dynamics of the phase transition discussed here.
}
\begin{align}
\kappa_{{\bf 15},{\bf 1}}=0 ,
\end{align}
and focus on the quartic coupling $\lambda_{{\bf 15},{\bf 1}}$.
With this mixing interaction,
the minimum in the potential $V_{\bf 15}(\Phi_{\bf 15})$ should not be a minimum in the full potential but be a saddle point.
Namely, it should be a minimum in the direction of $\Phi_{\bf 15}$ but a maximum in the direction of $\Phi_{\bf 1}$,
so that the intermediate phase with non-zero $\Phi_{\bf 15}$ (red point in Fig.~\ref{fig:two-step}) can terminate and the transition into the true minimum (blue point) safely happens at sufficiently low temperature.
In addition, this mixing interaction should not destroy the minimum in the potential $V_{\bf 1}(\Phi_{\bf 1})$ but should make it an absolute minimum.
It turns out that the condition to realize this vacuum structure is given by
\begin{align}
    2 \mu_\textbf{15}^2 < \lambda_{\textbf{15},\textbf{1}} v_{\bf 1}^2 < 2 \frac{\lambda_{\bf 1}\lambda_{\textbf{15} 1} v_{\bf 1}^4}{\mu_\textbf{15}^2}
\end{align}
in the tree-level analysis,
where $v_{\bf 1}=
\mu_{\bf 1}/\sqrt{\lambda_{\bf 1}} .$
Such a typical shape of the potential is shown in the bottom-right panel of Fig.~\ref{fig:VT_cont},
where $s$ and $h$ are proportional to $|\Phi_{\bf 1}|$ and $|\Phi_{\bf 15}|$, respectively.
(The precise definitions of $s$ and $h$ are given later.)

As for the Yukawa interactions,
we are only interested in those involving $\Phi_{\bf 15}$ and $\Phi_{\bf 1}$ to study effects on the phase transitions.
In particular, we focus on the interaction with the singlet fermions, 
$(y)_{ij}\,
\Psi_{\bf 1}^{(i)}
\Psi_{\bf 1}^{(j)}
\Phi_{\bf 1}$
and ignore that with the SM fermions
$(y_{15})_{\alpha\beta}\,
(\Phi_{\bf 15})_{ab}
(\Psi_{\bf \overline5}^{(\alpha)})^a
(\Psi_{\bf \overline5}^{(\beta)})^b$
(i.e., $(y_{15})_{\alpha\beta}$ is sufficiently small).
For simplicity, we further assume $(y)_{ij}$ to be proportional to the identity matrix, $(y)_{ij}=: y \delta_{ij}$.

To summarize, we have nine parameters in the relevant sector of the model,
\begin{align}
    N_f, \quad v_{\bf 1}, \quad g, \quad \mu_{\bf 15}, \quad \lambda_{\bf 1}, \quad \lambda_{\bm{15}1}, \quad \lambda_{\bm{15}2},\quad \lambda_{{\bf 15},{\bf 1}}, \quad y \,,
\end{align}
defined in Eqs.~\eqref{eq:yukawa}, \eqref{Eq:Potential_V15}, \eqref{Eq:Potential_V1}, and \eqref{Eq:Potential_Vmix-15+1}.
(We have replaced $\mu_{\bf 1}$ with $v_{\bf 1}$.)
We choose them for the four benchmark cases shown in Tab.~\ref{tab:BP-parameters}.

\begin{table}[tbp]
    \centering
     \begin{tabular}{|c|c|c|}
    \hline
        & $N_f$ &$v_{\bf 1}$ \\ \hline \hline
        BP1&$6$ & $10^{5} \,\mathrm{GeV} $\\ \hline
        BP2& $12$ & $10^{5} \,\mathrm{GeV} $\\ \hline
        BP3& $6$ & $10^{12} \,\mathrm{GeV} $\\ \hline
        BP4& $12$ & $10^{12} \,\mathrm{GeV} $\\ \hline
    \end{tabular}
 \hspace{1em}
    \begin{tabular}{|c|c|c|c|c|c|c|}
    \hline
    $g$& $\mu_{\bf 15}^2/v_{\bf 1}^2$ &$\lambda_{\bf 1}$ & $\lambda_{\bm{15}1}$ & $\lambda_{\bm{15}2}$ & $\lambda_{{\bf 15},{\bf 1}}$ & $y$  \\ \hline 
       $0.7$& $0.4$ & $0.05$ & $5.0$ & $1.0$ & $1.4$ & $ 0.6$ \\ \hline
    \end{tabular}
    \caption{
Benchmark parameter (BP) sets used in the analysis. 
We consider four BPs (BP1-4),
which have different $N_f$ and singlet VEV $v_{\bf 1}$ summarized in the left table and share the other parameters in the right table.
    }
    \label{tab:BP-parameters}
\end{table}

\section{Scenario}
\label{sec:scenario}
In this section, we present our scenario to solve the monopole problem.
This scenario involves three phase transitions (besides those in the SM),
which successively break and restore the gauge symmetries as follows:
\begin{align}
    SU(5) \to G_\mr{SM}\to SO(3)_C\times SO(2)_L\to G_\mr{SM},
\end{align}
where $G_\mr{SM}= SU(3)_C\times SU(2)_L \times U(1)_Y$.
Each phase transition is explained in more detail below.

\subsection{$SU(5)$ breaking: $U(1)_Y$ monopole production}
When the adjoint scalar $\Phi_{\bf 24}$ acquires a VEV  $v_{\bf 24}\simeq 10^{16}\, \mr{GeV}$, the $SU(5)$ GUT  gauge symmetry is spontaneously broken down to $G_\mr{SM}$.
This transition produces magnetic monopoles,
as indicated by the non-trivial second homotopy group $\pi_2$:
\begin{align}
    \pi_2(SU(5)/G_\mr{SM})\simeq \pi_1 (U(1))\simeq \mathbb{Z} \, .
\end{align}
This monopole has a magnetic charge associated with the unbroken $U(1)_Y$ gauge field,
and is therefore referred to as a $U(1)_Y$ monopole.
Since it has a very heavy mass $m_M \sim 4\pi\, v_{\bf 24}/g$,
it may cause the monopole problem.

The lower bound of the monopole number density at production
can be estimated by noting that the correlation length
$\ell_\mr{cor}$ during the phase transition is bounded
from above by the Hubble length scale~\cite{Kibble:1976sj,Zurek:1985qw}.
\begin{align}
    n_M(T_{\bf 24})\sim (\ell_\mr{cor})^{-3}\gtrsim H(T_{\bf 24})^3 \, ,
\end{align}
where $n_M(T)$ and $H(T)$ are the number density and the Hubble parameter at temperature $T$, respectively, and $T_{\bf 24}$ denotes a critical temperature at which the phase transition occurs.

The produced monopoles experience pair annihilation due to the attractive Coulomb interaction,
leaving a relic number density~\cite{Zeldovich:1978wj,Preskill:1979zi,Vilenkin:2000jqa}
\begin{align}
    \frac{n_M}{s}\sim 10^{-12} \frac{m_M}{10^{16}\,\mr{GeV}} \, ,
\end{align}
with $s$ being the entropy density.
Without any mechanism to further reduce the monopole number density, 
they would dominate the total energy density of the universe at
\begin{align}
    T_d \sim 10^4 \, \mr{GeV} \, \left(\frac{m_M}{10^{16}\, \mr{GeV}}\right)^2 \, ,
\end{align}
known as the monopole problem.
It should be noted that the Langacker-Pi mechanism around the electroweak scale (more generally, $T\lesssim T_d$) is unlikely to work sufficiently~\cite{Gates:1992gd,Farris:1991rg,Holman:1992xs}
because it suffers from the monopole domination.

\subsection{Special symmetry breaking: $U(1)_Y$ monopole disappearance}
After the $SU(5)$ symmetry is broken by the non-zero VEV $v_{\bf 24}$,
the symmetric tensor scalar $\Phi_{\bf 15}$ can develop a VEV $v_{\bf 15}$ as the temperature $T$ decreases,
even if it does not develop a VEV at zero temperature.
This can be understood by considering the one-loop thermal effective potential for $\Phi_{\bf 15}$ and $\Phi_{\bf 1}$.
We parametrize the scalar fields as
\begin{align}
    \Phi_{\bf 15} = \sqrt{\frac{1}{10}} h(x) \,1_{5\times 5} , \quad
    \Phi_{\bf 1} = s(x) \, ,  \label{eq:scalar-decomp}
\end{align}
with $h$ and $s$ real functions.
The finite-temperature effective potential is given by
\begin{align}
 V_T(h,s;T) = V_0(h,s) + V_\mr{CW} (h,s) + \Delta V_T (h,s;T) + \Delta V_\mr{ct} \label{eq:VT}
\end{align}
at the one-loop level,
where $V_0(h,s)$ is the tree-level potential obtained by substituting Eq.~\eqref{eq:scalar-decomp} into the potential in Eq.~(\ref{Eq:Potential_15+1}).
Regarding loop corrections, $V_\mr{CW} (h,s)$ is the Coleman-Weinberg potential~\cite{Coleman:1973jx}, and $\Delta V_T$ is the one-loop finite-temperature correction~\cite{Dolan:1973qd,Quiros:1999jp}.
The last term $\Delta V_\mr{ct}$ denotes finite counterterms to satisfy certain renormalization conditions, which will be specified in Sec.~\ref{sec:GW}.

The dominant contributions to $V_T$ are
given by the gauge bosons that acquire masses through the coupling with $\Phi_{\bf 15}$ (eight gauge bosons corresponding to $G_\mr{SM}\to SO(3)_C\times SO(2)_L$)
and the $N_f$ singlet fermions coupled with $\Phi_{\bf 1}$,
which are explicitly written within the high-temperature approximation as
\begin{align}
    V_T \sim \left(-\frac{\mu_\textbf{15}^2}{4} + \frac{2}{5} g^2 T^2 \right)h^2  
    +\left( -\frac{\mu_\textbf{1}^2}{2} + \frac{N_f}{12}y^2 T^2\right) s^2 
    + \cdots ,\label{eq:high-T}
\end{align}
where ``$\cdots$'' indicates cubic and quartic terms and other sub-dominant terms.
(The detailed form of the potential is given in the next section.)
From this expression, one finds that when 
\begin{align}
    \mu_\textbf{15}^2  > \frac {48 \mu_\textbf{1}^2 g^2}{5 N_f y^2} \label{eq:cond-of-mu15}
\end{align}
is satisfied,
there exists a temperature $T$ such that the coefficient of $h^2$ becomes negative while that of $s^2$ remains positive.
As a result, $\Phi_{\bf 15}$ ($h$) acquires a non-zero VEV while $\Phi_{\bf 1}$ ($s$) remains zero.
Such a non-trivial minimum appears around
\begin{align}
    T\sim \sqrt{\frac{5}{8}}\frac{\mu_\textbf{15}}{g} =: T_\textbf{15}\, . \label{eq:15-breaking-T}
\end{align}
The VEV is temperature-dependent and is approximately given by
\begin{align}
v_\textbf{15}(T) \sim \sqrt{\frac{10\mu_\textbf{15}^2-16 g^2 T^2}{5 \lambda_{\textbf{15}1}+\lambda_{\textbf{15}2}}}.
\end{align}
This VEV breaks $G_\mr{SM}$ further down to its special subgroup 
$SO(3)_C\times SO(2)_L$
(called special symmetry breaking).
Note that $\mu_\textbf{15}$, and hence $v_\textbf{15}(T)$, are much smaller than $v_\textbf{24}$, so that the $SU(5)$ breaking induced by $\Phi_{\bf 24}$ is not significantly affected.

Let us consider topological defects produced by this symmetry breaking.
First, $U(1)_Y$ is spontaneously broken to nothing,
leading to the production of a cosmic string associated with the non-trivial first homotopy group $\pi_1$,
\begin{align}
    \pi_1 (U(1)_Y) \simeq \mathbb{Z} \, .
\end{align}
This $U(1)_Y$ string connects a monopole and an antimonopole, forming a segment.
The string tension exerts an attractive force on the monopole--antimonopole pair, accelerating them toward each other.
Since the monopole core size is much smaller than the string width, the encounter of the monopole and antimonopole does not necessarily lead to immediate annihilation. 
They may overshoot each other and continue their motion, until the string tension pulls them back again. 
The pair therefore undergoes oscillatory motion, with a period set by the initial separation and the acceleration due to the string tension. 
This motion is eventually damped by friction from interactions with the thermal plasma.
For a monopole moving with velocity $\mathrm{v}$, the frictional force is estimated as
\begin{align}
   F_\mr{fric}\sim - g_* T^2 \,\mathrm{v } \label{eq:EOM}
\end{align}
in the non-relativistic limit, $\mathrm{v}\ll 1$.\footnote{
Scattering of the monopole off the string zero modes can provide an additional source of dissipation~\cite{Chitose:2025qyt}, whose magnitude is expected to be comparable to the frictional force in Eq.~\eqref{eq:EOM}.
}
One can estimate a relaxation time due to this friction, 
corresponding to the time scale to damp the oscillation,
as~\cite{Holman:1992xs,Gates:1992gd,Vilenkin:2000jqa}
\begin{align}
    \tau_\mr{rel}\sim \frac{m_M}{g_* T^2} \, ,
\end{align}
which is much shorter than the Hubble time scale since
\begin{align}
q\equiv\frac{\tau_\mr{rel}}{t} \sim \frac{m_M}{M_\mr{pl}}\sqrt{\frac{8\pi^3}{90g_*}} \ll 1 \, .
\end{align}

Since the string tension is almost constant for $T\lesssim T_\textbf{15}$,
the energy loss due to this dissipation decreases the separation $d$ (averaged over the oscillation) between the monopole-antimonopole pair:
\begin{align}
   \frac{\dot{d}}{d} \sim \frac{1}{\tau_\mr{rel}}
\Rightarrow    d\propto t^{-1/q}
\end{align}
from which we can conclude that the pair annihilates very quickly.

This phase transition also produces two additional types of monopoles,
namely $U(1)_L$ and $(\mathbb{Z}_2)_C$ monopoles.
The $U(1)_L$ monopole arises from the symmetry breaking 
$SU(2)_L \to SO(2)_L \simeq U(1)_L$ 
and corresponds to the standard 't Hooft--Polyakov monopole.
The $(\mathbb{Z}_2)_C$ monopole is more subtle.
Focusing on the color sector, the symmetry breaking can be written as
\begin{align}
    SU(3)_C\to SO(3)_C \simeq SU(2)/\mathbb{Z}_2 \, ,
\end{align}
which leads to the non-trivial second homotopy group
\begin{align}
    \pi_2(SU(3)_C/SO(3)_C)\simeq \pi_1(SU(2)/\mathbb{Z}_2)\simeq \pi_0(\mathbb{Z}_2) \simeq \mathbb{Z}_2 \, .
\end{align}
Thus it has a non-trivial winding in $SO(3)_C$ characterized by $\mathbb{Z}_2$~\cite{Weinberg:1983bf},
in the sense that a configuration with two monopoles is topologically the same as the vacuum configuration.
Neither the $U(1)_L$ nor $(\mathbb{Z}_2)_C$ monopoles cause the monopole problem because they will disappear as soon as the gauge symmetry is restored to $G_\mr{SM}$,
as discussed below.

\subsection{Symmetry restoration to $G_\mr{SM}$}
The gauge symmetry $SO(3)_C\times SO(2)_L$ is not compatible with the conventional late-time cosmology, and therefore it is necessary to restore the SM gauge group $G_\mr{SM}$ well before Big-Bang nucleosynthesis.
This requires that the VEV $v_\textbf{15}$ vanishes at a later stage, leading to the symmetry restoration 
$SO(3)_C\times SO(2)_L\to G_\mr{SM}$.

The symmetry restoration can be realized through a transition between two minima,
(i) $(h,s)=(v_\textbf{15}(T),0)$ and 
(ii) $(h,s)=(0,v_{\bf 1}(T))$,
where $v_{\bf 1}(T)$ is a temperature-dependent VEV of the singlet scalar, with $v_{\bf 1}(0)=v_{\bf 1}$.
(See Fig.~\ref{fig:two-step}.)
The second minimum appears at
\begin{align}
    T\sim \sqrt{\frac{6}{N_f}}\frac{\mu_{\bf 1}}{y} \, ,
\end{align}
which is smaller than Eq.~\eqref{eq:15-breaking-T} due to the condition given in  Eq.~\eqref{eq:cond-of-mu15}.
The transition from (i) to (ii) can be an FOPT depending on the model parameters.

\begin{figure}[tbp]
    \centering
    \includegraphics[width=0.4\textwidth]{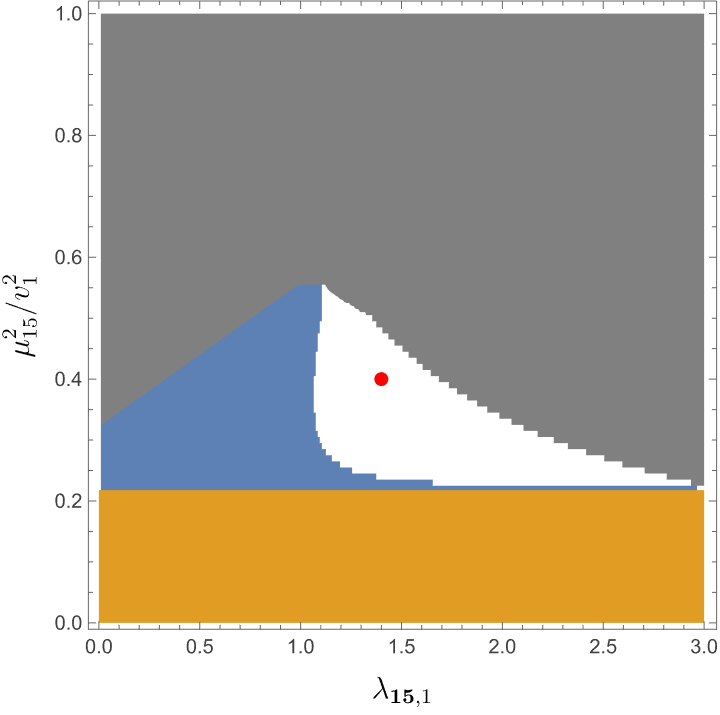}
    \hspace{5mm}
    \includegraphics[width=0.4\textwidth]{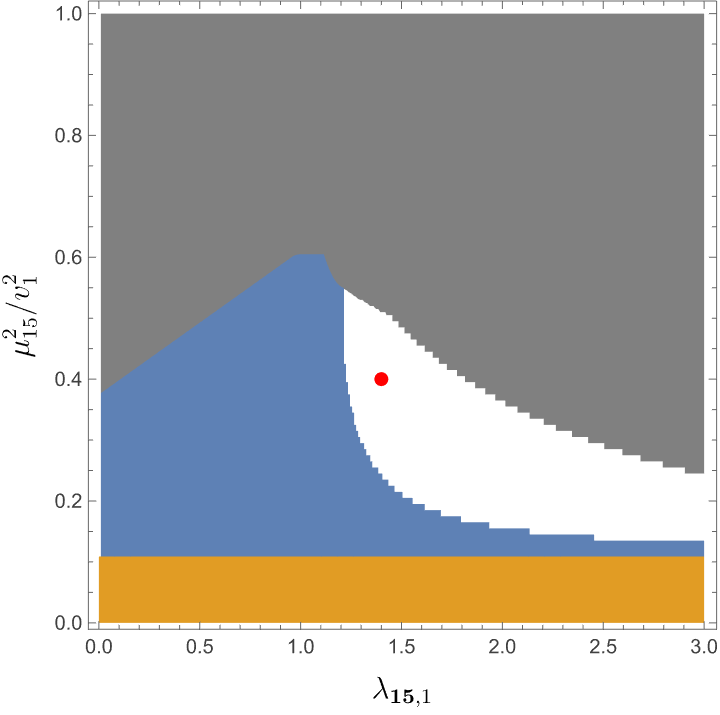}
    \caption{
Parameter space realizing the present scenario.
The gray region is excluded due to the absence of the correct vacuum structure at $T=0$.
The blue region corresponds to a second-order phase transition (SOPT) or a crossover for the symmetry restoration to $G_\mr{SM}$.
The orange region indicates that the intermediate special symmetry breaking ($h\neq 0$) does not occur, and hence the Langacker-Pi mechanism does not work.
The white region corresponds to the parameter space where the symmetry restoration proceeds via FOPT.
All parameters except for $\lambda_{\textbf{15},{\bf 1}}$ and $\mu_\textbf{15}$ are fixed to the benchmark values in Table~\ref{tab:BP-parameters}, with $N_f=6$ (left) and $N_f=12$ (right).
The red dots denote the benchmark points listed in Table~\ref{tab:BP-parameters}.
}
    \label{fig:param_cond}
\end{figure}

Within the high-$T$ approximation \eqref{eq:high-T}, 
one can derive the condition under which the transition is first-order.
Let us consider a temperature $T_c$ at which (i) and (ii) are degenerate,
\begin{align}
    T_c = \sqrt{
    \frac{ \sqrt{5\lambda_{\bf 1} v_{\bf 1}^4 (5\lambda_{\textbf{15}1}+\lambda_{\textbf{15}2})}-5\mu_\textbf{15}^2}
    { \frac{N_f y^2}{6\lambda_{\bf 1}} \sqrt{5\lambda_{\bf 1}(5\lambda_{\textbf{15}1}+\lambda_{\textbf{15}2})} -8g^2}
    } \, .
\end{align}
The transition is first-order when 
\begin{align}
\left. \partial_{s,s}V_T(v_\textbf{15}(T_c),s;T_c)\right|_{s=0} >0 \, ,
\end{align}
which implies the presence of a potential barrier between the two vacua.
Figure~\ref{fig:param_cond} shows the parameter region in which the symmetry restoration occurs via a FOPT. 
The white region corresponds to the parameter space consistent with the present scenario, and the red dots denote the benchmark points in Table~\ref{tab:BP-parameters}.
Such a FOPT can generate stochastic GWs through bubble nucleation and expansion.

After the symmetry is restored, the $U(1)_L$ and $(\mathbb{Z}_2)_C$ monopoles disappear because their characteristic length scales are inversely proportional to $v_\textbf{15}$.
As $v_\textbf{15}$ vanishes, these configurations become unstable and decay into the vacuum.
Therefore, they do not contribute to the energy density of the Universe, and the subsequent cosmological evolution proceeds as in the standard scenario.
We thus conclude that the monopole problem is avoided in this setup without invoking inflation.

Note that the singlet fermions $\Psi_{\bf{1}}^{(i)}$ acquire their masses in this phase through the singlet VEV $v_{\bf1}$ and could be abundant if they are stable.
Thanks to the Yukawa couplings $(y_5)_{i\alpha}$ in Eq.~\eqref{eq:yukawa},
they can decay into SM light particles safely, so that we are left with the standard cosmology after this transition.

\section{Gravitational wave spectrum}
\label{sec:GW}

A key feature of the scenario is that $G_\mr{SM}$ is temporarily broken to $SO(3)_C\times SO(2)_L$ and then restored to $G_\mr{SM}$.
As discussed above, the restoration transition can be  an FOPT for appropriate choices of the parameters in the model.
In this section, we estimate the stochastic GW signal generated by this FOPT.
In contrast to the high-temperature approximation adopted in Sec.~\ref{sec:scenario}, we use the one-loop finite-temperature effective potential and retain the full thermal functions.
The calculation follows the standard Coleman--Weinberg and thermal-field-theory treatment~\cite{Coleman:1973jx,Dolan:1973qd,Quiros:1999jp}
and utilizes the GW spectrum templates for the sound-wave source obtained by lattice simulations~\cite{Hindmarsh:2013xza,Hindmarsh:2015qta,Caprini:2015zlo,Hindmarsh:2017gnf,Ellis:2018mja,Caprini:2019egz}.

Returning to the finite-temperature potential~\eqref{eq:VT}, 
with $h$ and $s$ defined in Eq.~\eqref{eq:scalar-decomp}, 
the one-loop Coleman--Weinberg contribution is given by~\cite{Coleman:1973jx}
\begin{align}
    V_\mr{CW} (h,s)  = \sum_{i}\frac{n_i}{64\pi^2} m_i(h,s)^4 \left(\log\frac{m_i(h,s)^2}{\mu^2}-C_i\right) ,
    \label{eq:VCW_sec5}
\end{align}
where
$i$ runs over species of particles appearing in the one-loop diagrams,
i.e., 
two mixed states arising from $h$ and $s$,
eight gauge bosons,
and $N_f$ singlet fermions $\Psi_{\bf 1}$,
while $m_i(h,s)$ represents the field-dependent mass,
explicitly given by
\begin{align}
    m_{\pm}^2 &= 
    \frac{1}{40}\left(-20\lambda_{\bf 1} v_{\bf 1}^2-10\mu_{\bf 15}^{2}+s^{2}(60\lambda_{\bf 1}+5\lambda_{\bf 15, 1})+h^{2}(3\tilde\lambda+5\lambda_{\bf 15, 1})\right) \nonumber \\
    & \hspace{0.5em}\pm \frac{1}{40}\Big[\left(-20\lambda_{\bf 1} v_{\bf 1}^2-10\mu_{\bf 15}^{2}+s^{2}(60\lambda_{\bf 1}+5\lambda_{\bf 15, 1})+h^{2}(3\tilde\lambda+5\lambda_{\bf 15, 1})\right)^{2} \nonumber \\
    &\hspace{2em}-20\Big(40\lambda_{\bf 1} v_{\bf 1}^2 \mu_{\bf 15}^{2}+s^{2}(-20\lambda_{\bf 1}\lambda_{\bf 15, 1} v_{\bf 1}^2-120\lambda_{\bf 1}\mu_{\bf 15}^{2})+60s^{4}\lambda_{\bf 1}\lambda_{\bf 15, 1} \nonumber \\
    & \hspace{3em}+h^{2}\left(-12\lambda_{\bf 1}v_{\bf 1}^2 \tilde\lambda-10\lambda_{\bf 15, 1}\mu_{\bf 15}^{2}+s^{2}\left(36\lambda_{\bf 1}\tilde\lambda-15\lambda_{\bf 15, 1}^{2}\right)\right) +3h^{4}\tilde\lambda \lambda_{\bf 15, 1}\Big)\Big]^\frac{1}{2}
\\
    m_{A}^2 &= \frac{2}{5} (g h)^2\\
    m_{\Psi_{\bf 1}}^2 &= (y s)^2 \,,
\end{align}
where $\tilde{\lambda}=5\lambda_{{\bf 15} 1}+\lambda_{{\bf 15}2}$
and $m_\pm$ denotes the heavier/lighter mass eigenvalues of the mixed states.
The constant $C_i$ is defined as
\begin{align}
    C_i=
    \begin{cases}
        \frac{3}{2} & \text{for scalars and fermions} \\
        \frac{5}{6} & \text{for gauge bosons}
    \end{cases} \, ,
\end{align}
and $n_i$ denotes the number of degrees of freedom (including a minus sign for fermions).
In the numerical analysis, we take $\mu = v_{\bf 1}$ as the reference renormalization scale.

The one-loop finite-temperature correction is given by~\cite{Dolan:1973qd,Quiros:1999jp}
\begin{align}
\Delta V_T(h,s;T)
&=
\sum_i\frac{n_i}{2\pi^2} T^4 J_{b/f}(m_i(h,s)/T),
\label{eq:thermal_sec5}
\end{align}
where the thermal functions $J_{b/f}$ correspond to bosonic and fermionic contributions, respectively,
\begin{align}
J_b(x)
&=
\mathrm{Re}
\int_0^\infty dy\, y^2
\ln\left[
1 - \exp\left(-\sqrt{y^2+x^2}\right)
\right],
\\
J_f(x)
&=
\mathrm{Re}
\int_0^\infty dy\, y^2
\ln\left[
1 + \exp\left(-\sqrt{y^2+x^2}\right)
\right] \, .
\end{align}
For bosonic zero Matsubara modes, the perturbative expansion can be improved by daisy resummation (e.g., Refs.~\cite{Arnold:1992rz,Parwani:1991gq}),
which we do not include in the present analysis. 
We expect that this does not qualitatively affect our results
because the thermal barrier separating the two minima is essentially given by the quartic term $h^2 s^2$ instead of cubic terms.
A full treatment including daisy resummation is left for future work.

Even after removing the UV divergences in $V_\mr{CW}$ using the $\overline{\mr{MS}}$ scheme, 
it is useful to introduce finite counterterms $\Delta V_\mr{ct}$ 
so that the tree-level vacuum structure is preserved against the zero-temperature one-loop corrections:
\begin{align}
    \Delta V_\mr{ct}(h,s) = -\frac{\delta \mu_\textbf{1}^2}{2}  s^2 + \frac{\delta \lambda_\textbf{1}}{2}  s^4  -\frac{\delta \mu_\textbf{15}^2}{2}  h^2 + \frac{\delta \lambda_\textbf{15}}{2}  h^4 .
    \label{eq:Vct}
\end{align}
The four coefficients are fixed by requiring that the positions and depths of the two relevant extrema of the tree-level potential $V_0$ remain unchanged under $V_\mathrm{CW}+\Delta V_\mr{ct}$:
\begin{align}
  \Delta V_\mr{ct}(0,v_{\bf 1}) + V_\mr{CW}(0,v_{\bf 1})&=0, \\
   \Delta V_\mr{ct}(v_\textbf{15},0) +V_\mr{CW}(v_\textbf{15},0) &=0,\\
  \partial_s \left[\Delta V_\mr{ct}(0,s) + V_\mr{CW}(0,s)\right]_{s=v_{\bf 1}} &=0,\\  
  \partial_h \left[\Delta V_\mr{ct}(h,0) + V_\mr{CW}(h,0)\right]_{h=v_\textbf{15}} &=0 \, .
\end{align}
These conditions uniquely determine the four coefficients in Eq.~\eqref{eq:Vct}.

\begin{figure}[tbp]
    \centering
    \includegraphics[width=0.4\textwidth]{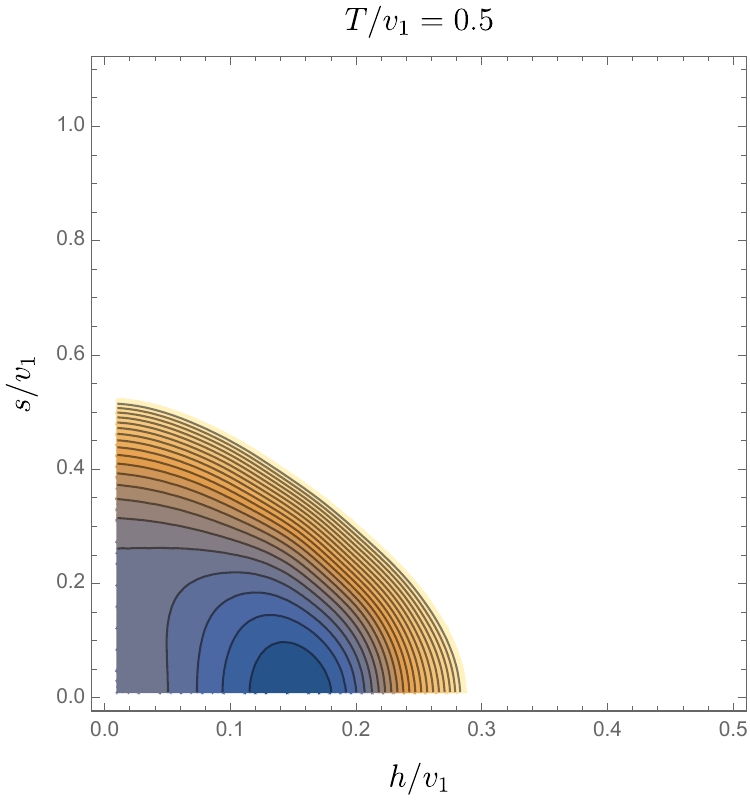}
    \hspace{2mm}
    \includegraphics[width=0.4\textwidth]{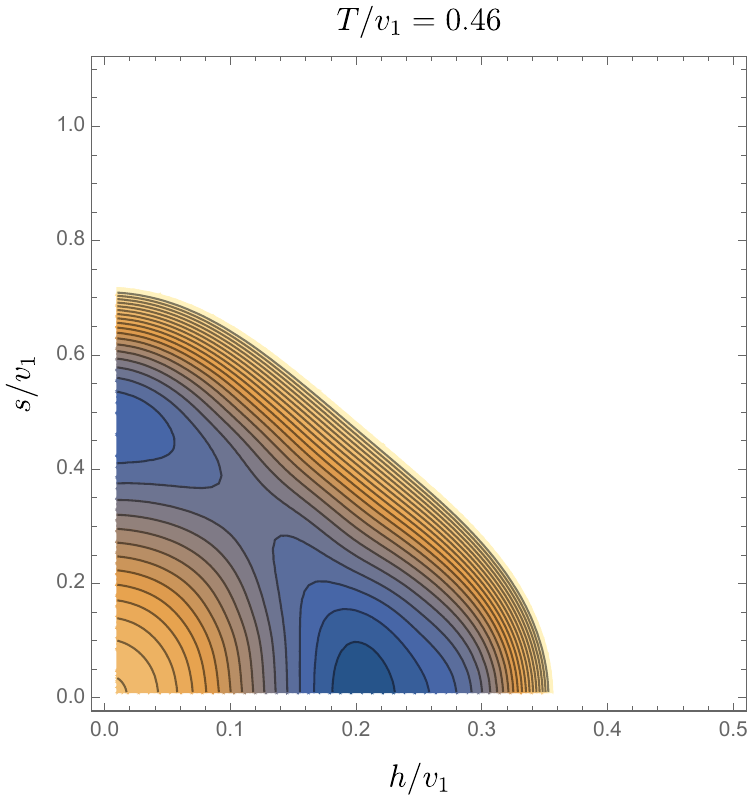}\\[2mm]
    \includegraphics[width=0.4\textwidth]{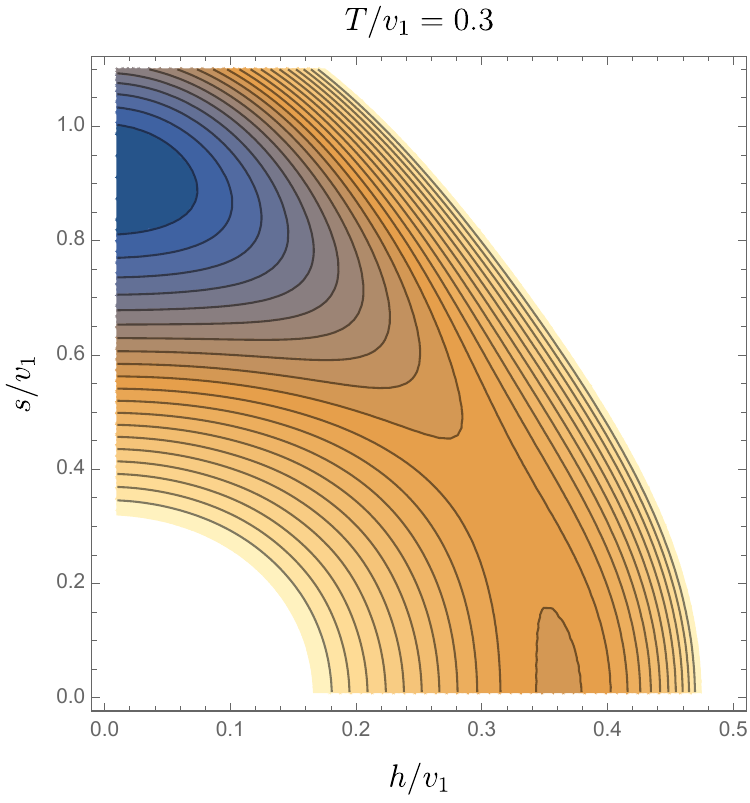}
    \hspace{2mm}
    \includegraphics[width=0.4\textwidth]{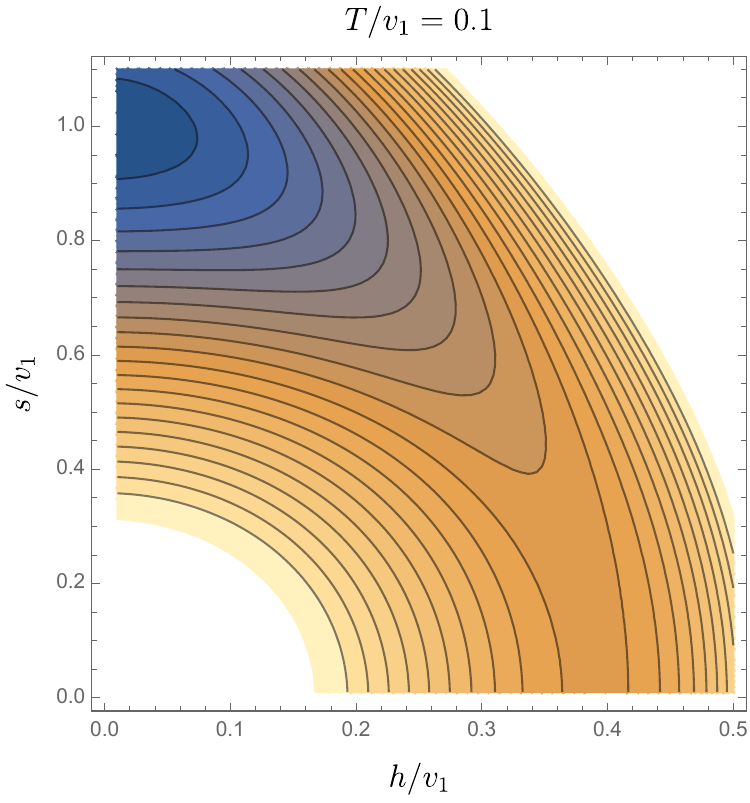}
\caption{Contour plots of the finite-temperature effective potential $V_T(h,s)$.
As the temperature decreases, a minimum with $s=0$ and $h\neq 0$ first appears, followed by the emergence of another minimum with $h=0$ and $s\neq 0$.
At lower temperatures, the latter becomes the true minimum, while the former disappears.
The parameters are taken to be those of BP3.
}
    \label{fig:VT_cont}
\end{figure}

The calculated potential $V_T(h,s)$ is shown in Fig.~\ref{fig:VT_cont},
in which the shaded colors correspond to its value.
At mildly high temperature,
there exists a minimum with $s=0$ and $h=v_{\bf 15}(T)\neq 0$ (top-left panel), 
followed by the emergence of another minimum with $h=0$ and $s=v_{\bf 1}(T)\neq 0$ (top-right and bottom-left panels).
At lower temperatures, the latter becomes the true minimum, while the former becomes a saddle point (bottom-right panel).

We now consider the GW signal produced during the FOPT.
Bubble nucleation, expansion, collision, and merger are known to source GWs, as established in early studies of cosmological phase transitions~\cite{Turner:1990rc,Kosowsky:1991ua,Kosowsky:1992vn,Kosowsky:1992rz,Kamionkowski:1993fg}.
For non-runaway transitions in a thermal plasma, numerical simulations show that the dominant contribution comes from the acoustic motion of the plasma~\cite{Hindmarsh:2013xza,Hindmarsh:2015qta,Hindmarsh:2017gnf}.
We therefore adopt the sound-wave contribution as our main estimate.

The relevant parameters of the phase transition are the strength parameter $\alpha$, the inverse duration parameter $\beta/H_*$, the wall velocity $v_w$, and the temperature $T_*$ at which the GWs are produced.
We define the energy difference between the false and true minima as
\begin{align}
    \Delta V(T)
    \equiv
    V_T\bigl(v_\textbf{15}(T),0;T\bigr)
    -
    V_T\bigl(0,v_{\bf 1}(T);T\bigr) .
    \label{eq:DeltaV_sec5}
\end{align}
Here the false vacuum corresponds to the intermediate phase with $h\neq 0$, while the true vacuum after the transition has $s\neq 0$.
The strength parameter is defined as
\begin{align}
\alpha
=
\frac{1}{\rho_{\rm rad}}
\left[
\Delta V
-
T \frac{d \Delta V}{dT}
\right]_{T=T_*},
\qquad
\rho_{\rm rad}=\frac{\pi^2}{30}g_*T_*^4  \, ,
\label{eq:alpha_sec5}
\end{align}
where $g_*$ is the effective number of relativistic degrees of freedom at $T=T_*$.

We define the nucleation temperature $T_n$ as the temperature at which the nucleation probability within one Hubble volume is of order unity,
\begin{align}
    \Gamma(T_n) \simeq H(T_n)^4 .
    \label{eq:nucleation}
\end{align}
The nucleation rate is written as $\Gamma\simeq A(T)\exp[-S_3(T)/T]$~\cite{Coleman:1977py,Linde:1981zj}, with $A(T)\sim T^4$,
and $S_3(T)$ being the three-dimensional Euclidean bounce action
\begin{align}
S_3[T]
=
4\pi
\int_0^\infty dr\, r^2
\left[
\frac{1}{2}
\left(
\frac{d h_c(r)}{dr}
\right)^2
+\frac{1}{2}
\left(
\frac{d s_c(r)}{dr}
\right)^2
+
V_T(h_c,s_c;T)-V_T(v_{\bf 15}(T),0;T)
\right],
\label{eq:S3_sec5}
\end{align}
where $(h_c(r),s_c(r))$ is the bounce solution.
This condition implies approximately
\begin{align}
   \frac{S_3(T_n)}{T_n}
   \simeq
   4\log\frac{T_n}{H(T_n)}
   +\cdots \, ,
\label{eq:nucleation_condition_sec5}
\end{align}
where the dots denote the mild dependence on the prefactor.

The inverse duration parameter is defined as
\begin{align}
\frac{\beta}{H_*}
\simeq 
T
\frac{d}{dT}
\left(
\frac{S_3}{T}
\right)
\bigg|_{T=T_n} .
\label{eq:beta_sec5}
\end{align}
A more precise treatment uses the percolation temperature $T_p$, defined in terms of the false-vacuum volume fraction, rather than $T_n$.
In this work we focus on benchmark points with $\alpha\lesssim0.1$,
for which the transitions are not strongly supercooled.
We therefore use $T_n$ as a proxy for $T_p$ and $T_*$.

\begin{table}[tbp]
    \centering
    \begin{tabular}{c|c|c|c}
 & $T_n/v_{\bf 1}$  & $\beta/H_*$ & $ \alpha $ \\ \hline \hline
        BP1& $0.23$ & $8.7\times 10^1$ & $4.4\times 10^{-2}$ \\ \hline
        BP2& $0.18$ & $3.0\times 10^3$ & $1.1\times 10^{-1}$ \\ \hline
        BP3& $0.19$ & $1.0\times 10^2$ & $7.9\times 10^{-2}$ \\ \hline
        BP4& $0.18$ & $2.7\times 10^4$ & $1.2\times 10^{-1}$ \\ \hline
    \end{tabular}
    \caption{
    Phase-transition parameters for the benchmark points used in the GW analysis.
    Here $T_n$ is the nucleation temperature and is used as a proxy for $T_*$.
    }
    \label{tab:BP-observables}
\end{table}

The sound-wave contribution to the present GW spectrum is estimated using the simple template assuming the single-peak broken-power law~\cite{Hindmarsh:2015qta,Hindmarsh:2017gnf,Caprini:2015zlo,Ellis:2018mja,Caprini:2019egz,Ellis:2020awk}
\begin{align}
\Omega_{\rm sw} h^2(f)
&=
2.65 \times 10^{-6}\,
(H_* \tau_{\rm sw})
\left(
\frac{\beta}{H_*}
\right)^{-1}
v_w
\left(
\frac{\kappa_v \alpha}{1+\alpha}
\right)^2
\left(
\frac{g_*}{100}
\right)^{-1/3}
\nonumber\\
&\quad \times
\left(
\frac{f}{f_{\rm sw}}
\right)^3
\left[
\frac{7}{4+3(f/f_{\rm sw})^2}
\right]^{7/2} .
\label{eq:Omega_sw}
\end{align}
Here $\tau_{\rm sw}$ denotes the lifetime of the acoustic source.
The peak frequency observed today is given by
\begin{align}
f_{\rm sw}
=
1.9 \times 10^{-5}\,
\frac{1}{v_w}
\left(
\frac{\beta}{H_*}
\right)
\left(
\frac{T_*}{100\,{\rm GeV}}
\right)
\left(
\frac{g_*}{100}
\right)^{1/6}
{\rm Hz} .
\label{eq:f_sw}
\end{align}
The finite lifetime of the sound-wave source is taken into account as~\cite{Hindmarsh:2017gnf,Ellis:2018mja,Caprini:2019egz,Ellis:2020awk}
\begin{align}
\tau_{\rm sw}
=
\min
\left[
\frac{1}{H_*},
\frac{R_*}{\bar{U}_f}
\right],
\qquad
H_*R_*
=
\max(v_w,c_s)
(8\pi)^{1/3}
\left(
\frac{\beta}{H_*}
\right)^{-1} ,
\label{eq:tausw_sec5}
\end{align}
where $c_s=1/\sqrt{3}$ and $R_*$ is the mean bubble separation.
The root-mean-square fluid velocity is approximated by~\cite{Hindmarsh:2015qta,Hindmarsh:2017gnf}
\begin{align}
\bar{U}_f^{\,2}
\simeq
\frac{3}{4}
\frac{\kappa_v\alpha}{1+\alpha} .
\end{align}
with $\kappa_v$ the efficiency factor.

The bubble-wall velocity is one of the largest theoretical uncertainties in the GW prediction.
A first-principles determination requires 
solving the scalar-field equation together with the plasma dynamics, either in local thermal equilibrium or including out-of-equilibrium effects through Boltzmann equations~\cite{Moore:1995si,Moore:1995ua,Laurent:2020gpg,Dorsch:2021nje,Laurent:2022jrs,DeCurtis:2022hlx,DeCurtis:2023hil,Dorsch:2023tss,Ekstedt:2024fyq,Dorsch:2024jjl,DeCurtis:2024hvh,Branchina:2025adj,Branchina:2025jou}.
In the absence of such calculations for the present model, we adopt $v_w=0.6$ as a benchmark estimate.
We emphasize that this may be a somewhat optimistic benchmark, since the actual wall velocity is controlled by plasma friction and is model dependent.
For the efficiency factor $\kappa_v$, 
as $c_s<v_w<v_J$ (with $v_J$ being the Jouguet velocity) in our benchmark points, 
we use the fitting formula of $\kappa_v$ for supersonic deflagrations given in Appendix of Ref.~\cite{Espinosa:2010hh}.
Although the single-broken-power-law fit in Eq.~\eqref{eq:Omega_sw} is sufficient for our purposes,
more refined sound-shell calculations can introduce a second spectral break associated with the thickness of the sound shell, which may modify the detailed shape around the peak~\cite{Hindmarsh:2016lnk,Hindmarsh:2019phv,RoperPol:2023dzg,Guo:2024wvp}.

\begin{figure}[tbp]
    \centering
    \includegraphics[width=0.7\textwidth]{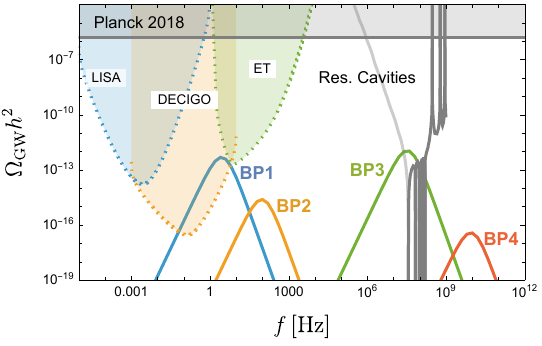}
    \caption{Stochastic GW spectrum produced by the first-order symmetry-restoration transition in our scenario.
    The spectra are computed using the sound-wave template in Eq.~\eqref{eq:Omega_sw}.
    The four colored peaks correspond to the benchmark points given in Table~\ref{tab:BP-parameters}.
    }
    \label{fig:GW_spectrum}
\end{figure}

The calculated phase-transition parameters and the resulting GW spectra for our benchmark points are shown in Tab.~\ref{tab:BP-observables} and Fig.~\ref{fig:GW_spectrum}.
The shaded regions in the figure are the projected sensitivity curves of LISA~\cite{LISA:2017pwj} (blue), DECIGO~\cite{DECIGO:2001} (orange), and the Einstein Telescope~\cite{EinsteinTelescope:2010} (green), taken from Ref.~\cite{Schmitz:2020syl}.
The gray horizontal band at the top is excluded by the bound on $N_\mathrm{eff}$ during Big-Bang Nucleosynthesis obtained from Planck observations~\cite{Planck:2018vyg}.
The gray spiky curve represents the projected sensitivity of a proposed resonant-cavity experiment~\cite{Herman:2022fau}.

The peak frequency scales approximately as $f_\mr{sw}\propto (\beta/H_*)T_*/v_w$.
Therefore, the benchmarks with $v_{\bf 1}=10^5\,{\rm GeV}$ can fall within the frequency range relevant for space- and ground-based interferometers, whereas those with $v_{\bf 1}=10^{12}\,{\rm GeV}$ are shifted to much higher frequencies and are more naturally compared with proposed high-frequency experiments such as resonant-cavity searches.
In particular, BP1 and BP3 may be probed by DECIGO and resonant-cavity experiments, and hence the symmetry-restoration transition $SO(3)_C\times SO(2)_L\to G_\mr{SM}$ in our scenario may be testable.

\section{Summary and discussions}

In this paper we have investigated the realization of the
Langacker--Pi mechanism to resolve the cosmological monopole
problem in an $SU(5)$ GUT framework in which the gauge symmetry
is broken into a special subgroup during the cosmological
evolution.
Our main focus has been on the vacuum structure of a scalar sector
containing the symmetric tensor scalar $\Phi_{\bf 15}$
and the singlet scalar $\Phi_{\bf 1}$.
We have analyzed the scalar potential including the mixing
interactions among these fields and examined the conditions under
which the symmetry-breaking sequence can proceed through an
intermediate phase in which the gauge symmetry is broken as
$SU(5)\rightarrow G_\mr{SM} \to SO(3)_C\times SO(2)_L$.
Since $SO(3)_C\times SO(2)_L$ is a special subgroup of $G_\mr{SM}$ (and hence $SU(5)$), the corresponding
vacuum does not support topologically stable monopoles.
Therefore, if such a phase appears during the cosmological
evolution, the monopole abundance generated in the initial
$SU(5)$ breaking can be efficiently reduced through the
Langacker--Pi mechanism.

We have also explored the possibility of the subsequent first-order symmetry-restoration transition back into $G_\mr{SM}$.
Such a phase transition may lead to the production of a
stochastic GW background which can in principle be tested by future GW experiments such as space- and ground-based interferometer and resonant cavity experiments.
The quantitative prediction of the GW spectrum, however, depends sensitively on the detailed
particle content and the thermal dynamics of the model,
and therefore a more realistic analysis is left for
future investigation.

Several phenomenological aspects that are usually considered in
realistic GUT models have not been addressed in the present work.
In particular, we have not attempted to construct realistic Yukawa
interactions for quarks and leptons, nor have we examined in
detail the conditions for gauge coupling unification or the
constraints arising from proton decay.
These issues depend sensitively on the detailed particle
content and heavy-field spectrum of the model and therefore
require a more complete model-building framework.
It would be interesting to extend the present setup by
incorporating realistic fermion mass structures, studying the
renormalization-group evolution of the gauge couplings, and
examining whether proton decay can be sufficiently suppressed
in such extended models.
We hope that the present work provides a useful starting point for constructing more realistic GUT models in which the cosmological monopole problem can be addressed through symmetry breaking into special subgroups during the cosmological evolution.
In addition, it may be interesting to investigate the possibility of baryogenesis through the monopole annihilation in our scenario~\cite{Davis:1992ca},
as well as primordial magnetogenesis~\cite{Turner:1987bw,Ratra:1991bn,Vachaspati:1991nm},
where the $U(1)_Y$ string segments ending on monopole–antimonopole pairs may play a role analogous to electroweak dumbbells~\cite{Vachaspati:1994xc,Urrestilla:2001dd,Patel:2023sfm,Patel:2023ybi},
although the efficiency of such a mechanism is highly nontrivial since the $U(1)_Y$ gauge field is massive during the intermediate phase in our scenario.

\section*{Acknowledgments}
The authors would like to thank
Akifumi Chitose
and
Thomas Konstandin
for useful discussions,
as well as Maaz Hayat for pointing out typographical errors.
This work is supported by the Deutsche Forschungsgemeinschaft under Germany's Excellence Strategy - EXC 2121 Quantum Universe - 390833306 (Y.H.)
and by the Japan Society for the Promotion of Science KAKENHI Grant
No.~JP21H05182 (N.Y.) and 
No.~JP26K17155 (Y.H.).


\appendix

\section{Symmetry breaking by individual scalar fields}
\label{Appendix:Vacuum}

In this appendix we briefly summarize the vacuum structures
that arise when each scalar multiplet is considered
independently.
The scalar potential of the full model is given in
Eq.~(\ref{Eq:Potential_Total}), while here we discuss the
vacuum structures associated with the individual scalar
fields for later reference.
General features of symmetry breaking by scalar fields in
$SU(N)$ gauge theories have been discussed in the literature
\cite{Li:1973mq}.

For $SU(n)$ gauge symmetry, spontaneous symmetry breaking
through the Higgs mechanism
\cite{Higgs:1964pj,Englert:1964et,Guralnik:1964eu}
depends on the representation
of the scalar field.
A nonvanishing VEV of a scalar
field in the fundamental representation ${\bf n}$ breaks
$SU(n)$ to $SU(n-1)$, while a VEV in the adjoint
representation ${\bf n^2-1}$ breaks $SU(n)$ to
$SU(m)\times SU(n-m)\times U(1)$.
Scalar fields in tensor representations lead to additional
patterns:
a second-rank symmetric tensor ${\bf n(n+1)/2}$ may break
$SU(n)$ to $SU(n-1)$ or $SO(n)$,
whereas a second-rank antisymmetric tensor
${\bf n(n-1)/2}$ may break $SU(n)$ to
$SU(n-2)$ or $USp(2\ell)$ ($\ell := [n/2]$).

The subgroups
$SU(n-1)$,
$SU(m)\times SU(n-m)\times U(1)$,
and $SU(n-2)$
are examples of {\it regular subgroups} of $SU(n)$,
while $SO(n)$ and $USp(2\ell)$ are
{\it special subgroups} (or irregular subgroups)
\cite{Dynkin:1957um,Dynkin:1957ek}.
A subgroup $H$ of a group $G$ is called a regular subgroup
if the Cartan subalgebra of $H$ is contained in that of $G$;
otherwise it is called a special subgroup.
Regular subgroups can be obtained by deleting nodes from
the (extended) Dynkin diagrams, whereas special subgroups
cannot be obtained in this way.
For reviews see, for example,
Refs.~\cite{Cahn:1985wk,Slansky:1981yr,Yamatsu:2015npn}.

When discussing spontaneous symmetry breaking it is useful
to consider the {\it little group} associated with a scalar
field configuration.
The little group $H_\phi$ of a vector $\phi$ in a
representation $\bm R$ of a group $G$ is defined by
\begin{align}
H_\phi := \left\{ g \in G \mid g\phi = \phi \right\}.
\label{eq}
\end{align}
The unbroken gauge symmetry after spontaneous symmetry
breaking corresponds to the little group of the VEV of the
scalar field.

In practice, the possible vacuum structures are often
strongly constrained by Michel's conjecture
\cite{Michel:1980pc}, which states that extrema of a
generic scalar potential constructed from a scalar field
in an irreducible representation tend to preserve one of
the maximal little groups of that representation.
This observation allows one to restrict the analysis to a
limited number of candidate symmetry-breaking patterns.

The VEVs of scalar fields are determined by minimizing the
scalar potential.
Denoting the potential evaluated at the VEVs by
$V({v_{\bf x}})$, the stationary conditions are
\begin{align}
\frac{\partial V}{\partial v_{\bf x}}=0 ,
\end{align}
which determine the extrema of the potential.
For each set of parameters these equations yield candidate
vacua, and the true vacuum corresponds to the configuration
with the lowest value of the potential.

\subsection{Adjoint scalar}
\label{Appendix:Vacuum-Adjoint}

We  analyze the vacuum structure of the adjoint scalar
$\Phi_{\bf 24}$ given in Eq.(\ref{Eq:Potential_V24}).
Since $\Phi_{\bf 24}$ is Hermitian, it can be diagonalized by an
$SU(5)$ transformation.
Therefore the vacuum expectation value can be written as
\begin{align}
\langle \Phi_{\bf 24} \rangle
=
{\rm diag}(v_1,v_2,v_3,v_4,v_5),
\end{align}
with the traceless condition
\begin{align}
v_1+v_2+v_3+v_4+v_5=0 .
\end{align}

The stationary conditions are obtained from
\begin{align}
\frac{\partial V_{\bf 24}}{\partial v_a}=0 .
\end{align}
The extrema correspond to several possible symmetry-breaking
patterns depending on the degeneracy of the eigenvalues.

The extrema correspond to several possible symmetry-breaking
patterns depending on the degeneracy of the eigenvalues.
According to Michel's conjecture, the extrema of a generic
potential tend to preserve one of the maximal little groups
of the representation.
For the adjoint representation of $SU(5)$, the maximal
little groups are
\begin{align}
SU(5)
\rightarrow
\left\{
\begin{array}{l}
SU(3) \times SU(2) \times U(1) \\
SU(4)\times U(1).\\
\end{array}
\right..
\end{align}

First we consider $SU(3)\times SU(2)\times U(1)$ breaking case.
The VEV corresponding to the SM
subgroup can be written in the normalized form
\begin{align}
\langle \Phi_{\bf 24} \rangle=
v_{\bf 24}
\frac{1}{\sqrt{60}}
{\rm diag}(2,2,2,-3,-3).
\label{Eq:VEV-24-5-321}
\end{align}
For this configuration the group invariants are
\begin{align}
{\rm Tr}(\Phi_{\bf 24}^2)
=\frac12 v_{\bf 24}^2,
\
{\rm Tr}(\Phi_{\bf 24}^3)
=
-\frac{1}{4\sqrt{15}}v_{\bf 24}^3,
\
{\rm Tr}(\Phi_{\bf 24}^4)
=
\frac{7}{120}v_{\bf 24}^4 .
\end{align}
Substituting these expressions into the scalar potential 
given in Eq.(\ref{Eq:Potential_V24})
gives
\begin{align}
V_{\bf 24}(v_{\bf 24})
&=
-\frac14\mu_{\bf 24}^2 v_{\bf 24}^2
-\frac{1}{12\sqrt{15}}\kappa_{\bf 24}v_{\bf 24}^3
+
\frac{1}{16}\lambda_{{\bf 24}1}v_{\bf 24}^4
+
\frac{7}{480}\lambda_{{\bf 24}2}v_{\bf 24}^4 .
\end{align}

Next we consider $SU(4)\times U(1)$ breaking case.
The VEV is given by
\begin{align}
\langle \Phi_{\bf 24} \rangle=
v'_{\bf 24}
\frac{1}{\sqrt{40}}
{\rm diag}(1,1,1,1,-4).
\label{Eq:VEV-24-5-41}
\end{align}
For this configuration the invariants become
\begin{align}
{\rm Tr}(\Phi_{\bf 24}^2)
=\frac12 v_{\bf 24}'^2,
\
{\rm Tr}(\Phi_{\bf 24}^3)
=
-\frac{3}{4\sqrt{10}}v_{\bf 24}'^3,
\
{\rm Tr}(\Phi_{\bf 24}^4)
=
\frac{13}{80}v_{\bf 24}'^4 .
\end{align}
The scalar potential becomes
\begin{align}
V_{\bf 24}(v_{\bf 24}')
&=
-\frac14\mu_{\bf 24}^2 v_{\bf 24}'^2
-\frac{1}{4\sqrt{10}}\kappa_{\bf 24}v_{\bf 24}'^3
+
\frac{1}{16}\lambda_{{\bf 24}1}v_{\bf 24}'^4
+
\frac{13}{320}\lambda_{{\bf 24}2}v_{\bf 24}'^4 .
\end{align}

The boundedness of the potential follows from the positivity
conditions of the quartic invariants constructed from
$\Phi_{\bf 24}$.
This requires the quartic couplings to satisfy
\begin{align}
\lambda_{{\bf 24}1}>0,
\qquad
\lambda_{{\bf 24}1}+\frac15\lambda_{{\bf 24}2}>0 .
\end{align}
In addition, spontaneous symmetry breaking requires
\begin{align}
\mu_{\bf 24}^2>0 .
\end{align}

Comparing the potential energies of the two extrema shows that the
vacuum aligned in the $(2,2,2,-3,-3)$ direction is energetically
favored in a broad region of parameter space when
\begin{align}
\lambda_{{\bf 24}2} > 0 ,
\end{align}
corresponding to the symmetry breaking $SU(5)\rightarrow SU(3)\times SU(2)\times U(1)$.
On the other hand, when $\lambda_{{\bf 24}2}$ becomes sufficiently
negative, the vacuum aligned in the $(1,1,1,1,-4)$ direction may
become the global minimum, corresponding to the symmetry breaking $SU(5)\rightarrow SU(4)\times U(1)$.

\subsection{Fundamental scalar}
\label{Appendix:Vacuum-Fundamental}

We next briefly discuss the vacuum structure of the
fundamental scalar field $\Phi_{\bf 5}$ given in Eq.~(\ref{Eq:Potential_V5})

The vacuum expectation value of $\Phi_{\bf 5}$ determines the
unbroken subgroup of $SU(5)$.
For a scalar field in the fundamental representation,
the maximal little group is $SU(4)$.
Accordingly the VEV can be rotated into the form
\begin{align}
\langle \Phi_{\bf 5} \rangle^T=
\frac{v_{\bf 5}}{\sqrt2}
\begin{pmatrix}
0&
0&
0&
0&
1\\
\end{pmatrix}.
\label{Eq:VEV-5-5-4}
\end{align}
Under an $SU(5)$ transformation
\begin{align}
\langle \Phi_{\bf 5} \rangle
\rightarrow
U
\langle \Phi_{\bf 5} \rangle .
\end{align}
The vacuum remains invariant if
\begin{align}
U_a^{\ 5}=\delta_{a5}.
\end{align}
This condition fixes the fifth component, while the
transformations acting on the first four components form an
$SU(4)$ subgroup.
Therefore the symmetry breaking pattern is
\begin{align}
SU(5)\rightarrow SU(4).
\end{align}

Substituting the VEV into the scalar potential given in Eq.(\ref{Eq:Potential_V5}),
the potential becomes
\begin{align}
V_{\bf 5}(v_{\bf 5})=
-\frac14 \mu_{\bf 5}^2 v_{\bf 5}^2
+
\frac{1}{16}\lambda_{\bf 5} v_{\bf 5}^4 .
\end{align}
Minimizing the potential with respect to $v_{\bf 5}$,
\begin{align}
\frac{\partial V_{\bf 5}}{\partial v_{\bf 5}}=0 ,
\end{align}
gives
\begin{align}
v_{\bf 5}^2=\frac{2\mu_{\bf 5}^2}{\lambda_{\bf 5}} .
\end{align}
The vacuum energy is
\begin{align}
V_{\bf 5}(v_{\bf 5})=
-\frac{\mu_{\bf 5}^4}{4\lambda_{\bf 5}} .
\end{align}

The stability of the potential at large field values requires
\begin{align}
\lambda_{\bf 5}>0 .
\end{align}
In addition, spontaneous symmetry breaking requires
\begin{align}
\mu_{\bf 5}^2>0 .
\end{align}

\subsection{Symmetric tensor scalar}
\label{Appendix:Vacuum-Tensor}

We analyze the vacuum structure of the symmetric tensor
scalar $\Phi_{\bf 15}$ given in Eq.(\ref{Eq:Potential_V15}).

The vacuum expectation value of $\Phi_{\bf 15}$ determines the
unbroken subgroup of $SU(5)$.
It is known that the symmetric tensor vacuum can lead to the
following maximal subgroups
\begin{align}
SU(5)\rightarrow
\begin{cases}
SU(4) & ({\rm regular})\\
SO(5) & ({\rm special})
\end{cases}
\end{align}
depending on the parameters of the scalar potential
\cite{Li:1973mq}.

We first consider the vacuum configuration
\begin{align}
\langle (\Phi_{\bf 15})_{ab} \rangle
=
\frac{v_\textbf{15}}{\sqrt{10}}\,
\delta_{ab}.
\label{Eq:VEV-15-5-5}
\end{align}
Under an $SU(5)$ transformation
\begin{align}
\langle (\Phi_{\bf 15})_{ab} \rangle
\rightarrow
\frac{v_\textbf{15}}{\sqrt{10}}\,
U_a^{\ c} U_b^{\ d} \delta_{cd}.
\end{align}
The vacuum remains invariant if
\begin{align}
U_a^{\ c}U_b^{\ d}\delta_{cd}=\delta_{ab},
\end{align}
which defines the subgroup $SO(5)$.
Therefore the symmetry breaking pattern is
\begin{align}
SU(5)\rightarrow SO(5).
\end{align}
Substituting the above vacuum configurations into the scalar
potential in Eq.(\ref{Eq:Potential_V15}), the potential can be
written as a function of the VEV $v_\textbf{15}$.
Minimizing the potential with respect to $v_\textbf{15}$,
\begin{align}
\frac{\partial V_{\bf 15}}{\partial v_\textbf{15}}=0,
\end{align}
determines the vacuum value of $v_\textbf{15}$.

Another possible vacuum alignment is
\begin{align}
\langle (\Phi_{\bf 15})_{ab} \rangle
=
\frac{v_\textbf{15}'}{\sqrt{2}}\,
\delta_{a5}\delta_{b5}.
\label{Eq:VEV-15-5-4}
\end{align}
Under an $SU(5)$ transformation
\begin{align}
\langle (\Phi_{\bf 15})_{ab} \rangle
\rightarrow
\frac{v_\textbf{15}'}{\sqrt{2}}\,
U_a^{\ c}U_b^{\ d}\delta_{c5}\delta_{d5}.
\end{align}
The vacuum remains invariant if
\begin{align}
U_a^{\ 5}=\delta_{a5}.
\end{align}
This condition fixes the fifth component,
while the transformations acting on the first four
components form an $SU(4)$ subgroup.
Thus the symmetry breaking pattern becomes
\begin{align}
SU(5)\rightarrow SU(4).
\end{align}

Substituting these vacuum configurations into the scalar
potential in Eq.(\ref{Eq:Potential_V15}),
we obtain the vacuum energies
\begin{align}
V_{\bf 15}(v_\textbf{15})
=
-\frac{5\mu_{\bf 15}^4}
{4(5\lambda_{{\bf 15}1}+\lambda_{{\bf 15}2})},
\quad
V_{\bf 15}(v_\textbf{15}')
=
-\frac{\mu_{\bf 15}^4}
{4(\lambda_{{\bf 15}1}+\lambda_{{\bf 15}2})}.
\end{align}

Comparing the vacuum energies shows that
\begin{align}
\lambda_{{\bf 15}2}>0
\label{Eq:Condition-SU5-to-SO5}
\end{align}
leads to the $SO(5)$ vacuum, while
\begin{align}
\lambda_{{\bf 15}2}<0
\label{Eq:Condition-SU5-to-SU4}
\end{align}
favors the $SU(4)$ vacuum.

The scalar potential contains two independent quartic
invariants,
$(\mathrm{Tr}\, (\Phi^\dagger\Phi))^2$
and
$\mathrm{Tr}(\Phi^\dagger\Phi\Phi^\dagger\Phi)$.
The difference between the two vacua originates from the
quartic invariant
${\rm Tr}(\Phi^\dagger\Phi\Phi^\dagger\Phi)$.
While both vacua satisfy
${\rm Tr}(\Phi^\dagger\Phi)=v_\textbf{15}^2/2$,
the value of
${\rm Tr}(\Phi^\dagger\Phi\Phi^\dagger\Phi)$
is larger for the $SU(4)$ vacuum than for the $SO(5)$ vacuum.
Therefore the sign of $\lambda_{{\bf 15}2}$ determines which
vacuum is energetically favored.

The stability of the scalar potential at large field values
requires
\begin{align}
\lambda_{{\bf 15}1}>0,
\qquad
\lambda_{{\bf 15}1}+\frac15\lambda_{{\bf 15}2}>0 .
\label{Eq:Condition-Stability}
\end{align}

In addition, spontaneous symmetry breaking requires
\begin{align}
\mu_{\bf 15}^2>0 .
\label{Eq:Condition-SSB}
\end{align}

\subsection{Singlet scalar}
\label{Appendix:Vacuum-Singlet}

We consider the singlet scalar field $\Phi_{\bf 1}$
with the potential given in Eq.~(\ref{Eq:Potential_V1}).

Since $\Phi_{\bf 1}$ is a singlet under $SU(5)$,
its VEV does not break the gauge
symmetry. We denote the VEV by
\begin{align}
\langle \Phi_{\bf 1} \rangle = v_{\bf 1} .
\end{align}
Substituting this into the scalar potential yields
\begin{align}
V_{\bf 1}(v_{\bf 1})=
-\frac12\mu_{\bf 1}^2 v_{\bf 1}^2
+
\frac13 \kappa_{\bf 1} v_{\bf 1}^3
+
\frac14\lambda_{\bf 1} v_{\bf 1}^4 .
\end{align}
The stationary condition
\begin{align}
\frac{\partial V_{\bf 1}}{\partial v_{\bf 1}}=0
\end{align}
gives the solutions
\begin{align}
v_{\bf 1}=0,\quad
v_{\bf 1}=
\frac{-\kappa_{\bf 1}\pm
\sqrt{\kappa_{\bf 1}^2+4\lambda_{\bf 1}\mu_{\bf 1}^2}}
{2\lambda_{\bf 1}} .
\end{align}

The stability of the potential requires
\begin{align}
\lambda_{\bf 1}>0 .
\end{align}
If $\mu_{\bf 1}^2>0$, the origin becomes unstable and a
nonzero VEV $v_{\bf 1}\neq0$ is generated.

We thus obtain the possible symmetry-breaking patterns
induced by each scalar multiplet individually.
These results provide a useful guide when analyzing
the vacuum structure of the full scalar potential.


\bibliographystyle{JHEP}
\bibliography{references}

\end{document}